\documentclass[draftcls,onecolumn,12pt]{IEEEtran}

\usepackage{cite}
\usepackage{bm}
\usepackage{url}
\usepackage[usenames,dvipsnames]{color}
\usepackage[dvipsnames]{xcolor}
\usepackage{amsmath}
\usepackage{mathtools}
\usepackage[]{amssymb}
\usepackage{amsthm}
\usepackage{float}
\usepackage{dsfont}
\usepackage{array}
\usepackage{booktabs}
\usepackage{multirow}
\usepackage{footmisc}

\usepackage[dvips]{graphicx}
\usepackage[para,flushleft]{threeparttable} 
\usepackage{multirow}	
\usepackage{dcolumn} 

\usepackage[ruled]{algorithm2e}
\usepackage{xifthen} 

\makeatletter
\newcommand{\removelatexerror}{\let\@latex@error\@gobble}
\makeatother

\newtheorem{theorem}{Theorem}
\newtheorem{lemma}[theorem]{Lemma}
\newtheorem{remark}{Remark}

\newif \ifwebcolor
\webcolortrue

\def \ba {\begin{array}}
\def \ea {\end{array}}
\def \benu {\begin{enumerate}}
\def \eenu {\end{enumerate}}
\def \bdes {\begin{description}}
\def \edes {\end{description}}

\def \bitem {\begin{itemize}}
\def \eitem {\end{itemize}}

\def \bfl {\begin{flushleft}}
\def \efl {\end{flushleft}}
\def \bfr {\begin{flushright}}
\def \efr {\end{flushright}}

\def \beq {\begin{equation}}
\def \eeq {\end{equation}}
\def \bqa {\begin{eqnarray}}
\def \eqa {\end{eqnarray}}
\def \bqa* {\begin{eqnarray*}}
\def \eqa* {\end{eqnarray*}}

\def \bal {\begin{align}}
\def \eal {\end{align}}

\newcounter{mytempeqncnt}

\DeclareMathOperator*{\argmax}{arg\,max}
\DeclareMathOperator*{\argmin}{arg\,min}

\begin{document}
%
\title{Clustered Sparse Channel Estimation for Massive MIMO Systems by 
Expectation Maximization-Propagation (EM-EP)}

\author{\IEEEauthorblockN{Mohammed Rashid, ~\IEEEmembership{Student Member,~IEEE}, Mort Naraghi-Pour, ~\IEEEmembership{Senior Member,~IEEE}}}

\maketitle
\thispagestyle{empty}
\pagestyle{empty}

\def\bda{\mathbf{a}}
\def\bdd{\mathbf{d}}
\def\bdg{\mathbf{g}} 
\def\bdh{\mathbf{h}}
\def\bdm{\mathbf{m}}
\def\bds{\mathbf{s}} 
\def\bdn{\mathbf{n}}
\def\bdp{\mathbf{p}}
\def\bdw{\mathbf{w}} 
\def\bdx{\mathbf{x}} 
\def\bdy{\mathbf{y}} 
\def\bdz{\mathbf{z}}
\def\bdA{\mathbf{A}}
\def\bdC{\mathbf{C}}
\def\bdD{\mathbf{D}} 
\def\bdF{\mathbf{F}}
\def\bdG{\mathbf{G}} 
\def\bdH{\mathbf{H}}
\def\bdI{\mathbf{I}}
\def\bdJ{\mathbf{J}}
\def\bdX{\mathbf{X}}
\def\bdK{\mathbf{K}}
\def\bdQ{\mathbf{Q}}
\def\bdR{\mathbf{R}}
\def\bdS{\mathbf{S}}
\def\bdV{\mathbf{V}}
\def\bdGamma{\bm{\Gamma}}
\def\bdgamma{\bm{\gamma}}
\def\bdalpha{\bm{\alpha}}
\def\bdmu{\bm{\mu}}
\def\bdSigma{\bm{\Sigma}}
\def\bdxi{\bm{\xi}}
\def\bdl{\bm{\ell}}
\def\bdLambda{\bm{\Lambda}}
\def\bdeta{\bm{\eta}}
\def\bdPhi{\bm{\Phi}}
\def\bdtheta{\bm{\theta}}
\def\btau{\bm{\tau}}
\def\deg{\text{o}}

\def\setA{\mathcal{A}} 
\def\bdsetS{\bm{\mathcal S}}
\def\bdL{{\mathcal L}}
\def\bdsetH{\bm{\mathcal H}}
\def\bdsetY{\bm{\mathcal Y}}
\def\famF{\mathcal{F}}
\def\tq{\tilde{q}}
\def\tbdJ{\tilde \bdJ}
\def\l{\ell}
\def\bdzero{\mathbf{0}} 
\def\Exp{\mathbb{E}} 
\def\exp{\text{exp}} 
\def\M{\mathcal{M}} 
\def\R{\mathbb{R}} 
\def\C{\mathbb{C}} 
\def\CN{\mathcal{CN}} 
\def\N{\mathcal{N}} 
\def\Es{E_s} 
\def\Bern{\text{Bern}}
\def\symset{\setA_\M} 
\def\symsetK{\symset^K} 
\begin{abstract}
We study the problem of downlink channel estimation in multi-user 
massive multiple input multiple output (MIMO) systems. To this end, we consider 
a Bayesian compressive sensing approach in which the clustered 
sparse structure of the channel in the angular domain is employed to reduce 
the pilot overhead. To capture the clustered structure, we employ a conditionally 
independent identically distributed Bernoulli-Gaussian prior on the sparse vector representing the 
channel, and a Markov prior on its support vector. An expectation propagation 
(EP) algorithm is developed to approximate the intractable joint 
distribution on the sparse vector and its support with a distribution from an 
exponential family. The approximated distribution is then used for direct 
estimation of the channel. The EP algorithm assumes that 
the model parameters are known a priori. Since these parameters are unknown, we 
estimate these parameters using the expectation maximization (EM) algorithm.
The combination of EM and EP referred to as EM-EP algorithm is reminiscent of the  
variational EM approach.
Simulation results
show that the proposed EM-EP algorithm outperforms several recently-proposed algorithms
in the literature.
\end{abstract}


\section{Introduction}



In FDD-based massive MIMO systems, downlink (DL) channel estimation is quite challenging \cite{effectv_dim3}.
In the conventional pilot-based method, the
length of the pilot sequence scales with the number of transmitting antennas.
This implies a long pilot sequence which results in reduced spectral efficiency. 
Moreover, the time required for pilot and data transmission may
exceed the coherence time of the channel.
Recently, compressive sensing (CS) \cite{BCS, CS} has been explored to reduce 
the pilot-overhead. 
Due to the limited local scattering in the propagation environment, massive MIMO 
channel has a sparse representation in the discrete Fourier transform (DFT) basis 
\cite{DFT_algo1, effectv_dim3, Bajwa_CS, A_Sayeed1}.
Using this sparsity structure, 
many CS-based estimation algorithms have been devised. 
The classical orthogonal matching pursuit (OMP) \cite{OMP_paper} and compressive sampling matching pursuit (CoSaMP) \cite{OMP_CoSaMP} are investigated in \cite{effectv_dim2, S_TCS_paper}. 
In \cite{effectv_dim1} the authors assumed a common spatial sparsity among the subcarriers in a frequency-selective DL channel and proposed the distributed sparsity adaptive matching pursuit (DSAMP). Using a similar common spatial sparsity assumption, 
a generalized approximate message passing (GAMP) based algorithm is proposed 
in \cite{DFT_algo4} and the sparse Bayesian learning (SBL) algorithm is derived in \cite{SBL_1,Tipping_2001}. 
These and other
algorithms which use a DFT basis to obtain the sparse representation, 
employ a fixed uniformly-spaced discrete grid in the angular domain which may not be  
sufficiently dense. 
As a result, some of the physical angles of departures (AoDs) of the massive MIMO channel may not lie on the assumed grid points. This direction mismatch error, also known as channel modeling error, causes leakage of energy from such physical AoDs into the nearby angular bins resulting in a straddle performance loss. In \cite{Overcomp_DFT1,Overcomp_DFT2} this modeling error is minimized by learning a 
better over-complete dictionary for the sparse representation. 
However, the proposed algorithm requires extensive
channel measurements from several locations 
in the cell to be used as training samples. These measurements are cell-specific and 
difficult to collect in practice. In \cite{off_grid1}, an Off-grid SBL algorithm is proposed in which 
the sampled grid points are modeled as continuous-valued parameters and are learned iteratively to reduce the modeling error. 
Simulation results in \cite{off_grid1} show improved performance of off-grid SBL compared to the over-complete dictionary 
learning algorithm in \cite{Overcomp_DFT1}.

SBL and off-grid SBL 
aim to recover the sparse vector coefficients 
individually by modeling them with an independent and identically distributed (iid) 
Gaussian prior distribution. 
However,
according to the geometry-based stochastic channel model (GSCM) 
\cite{GSCM_2003}, there are a few dominant 
scatterers in the propagation environment, and the sub-paths from each scatterer concentrate in small angular spreads 
which appear as non-zero clusters in the sparse representation.
This model is used in \cite{B_LASSO}, although with the 
stringent assumption of uniformly-sized clusters in the sparse vector. 
For non-uniform burst sparsity\footnote{This refers to the case when the non-zero clusters in the sparse vector appear with non-equal sizes separated with 
sequences of zeros of arbitrary length \cite{nU_Burst_Dai_2019}. \label{foot_1}}, a pattern-coupled SBL (PC-SBL) algorithm is proposed in \cite{PC_SBL1} in which the precision of each coefficient in the sparse vector is tuned according to 
the precision of its immediate neighbors. However, PC-SBL updates the precisions with a sub-optimal solution. 
To avoid this sub-optimality, a generic version of PC-SBL is 
derived in \cite{nU_Burst_Dai_2019}, referred to as PC-VB here, where 
the authors assigned a 
latent support vector to every coefficient in the sparse vector and assumed a multinoulli prior on the 
support vector. The resulting joint posterior 
distribution on the sparse vector and its support is approximated 
with a variational Bayes-based algorithm \cite{VB_algo}. Grid refining procedure 
from \cite{off_grid1} is also used to mitigate the direction mismatch errors. 
In the same vein, Turbo compressive sensing (TCS) algorithm and expectation maximization based GAMP algorithm were proposed in \cite{TCS_paper} and \cite{Vila_EM-BG-GAMP}, respectively, in which the sparse vector coefficients 
are modeled with an iid Bernoulli-Gaussian (BG) prior. 
In \cite{S_TCS_paper} the authors extended \cite{TCS_paper} for the 
clustered sparse structure of the massive MIMO channel and proposed a structured turbo compressive sensing (S-TCS) 
algorithm. With a conditional iid BG prior on the sparse vector, 
a Markov prior is assumed on its support to integrate the clustering 
information of the massive MIMO channel. 
In \cite{Super_Markov_1}, 
a super-resolution clustered sparse Bayesian learning (SuRe-CSBL) algorithm is 
proposed for a Markov prior distribution on the support vector. 
SuRe-CSBL approximates the true joint posterior distribution on the sparse 
vector and its support with a structured GAMP algorithm. The approximated distribution is then used for the  estimation of massive MIMO channel. 
{The grid refining method 
from \cite{off_grid1} is also integrated into SuRe-CSBL.}

In this paper, we propose an expectation propagation (EP) algorithm to estimate 
the clustered sparse vector 
representing the massive MIMO channel. Once the sparse vector is estimated from the 
received signal, the physical massive MIMO channel can be easily estimated by a transform operation on the 
sparse vector as in 
\cite{nU_Burst_Dai_2019,S_TCS_paper, Super_Markov_1}. 
The contributions made in this paper are summarized as follow:
\begin{itemize}

\item Expectation propagation (EP) algorithm \cite{Seeger_2005, Qi_2007} has been recently applied to SIMO and MIMO channel estimation \cite{Ghavami1, Ghavami2,
Ghavami3, Naraghi}. It has also been
applied to 
solve the inference problem in the CS literature \cite{CS_EP}. 
In \cite{Hernandez_2015}, the authors used an EP algorithm to approximate the true joint posterior distribution on 
the sparse vector and its support with a distribution from an exponential family. However, an iid Bernoulli prior 
is assumed on the support vector which does not capture the clustered structure of the sparse vector. 
{In \cite{Hernandez_2013} the authors assumed that the 
partitioning of the cluster in the sparse vector is known a priori and modeled each cluster with a different Bernoulli prior distribution. }
In contrast, we assume here that the cluster partitioning in the sparse vector 
is unknown. Therefore to capture the structure of the sparse vector we model its support vector with a first-order Markov process. An EP algorithm is developed to iteratively approximate the intractable true joint posterior distribution on 
the sparse vector and its support with a distribution from an exponential family. This 
distribution is then used for the direct estimation of the 
DL massive MIMO channel.  
  
\item The framework of EP algorithm in \cite{Hernandez_2013,Hernandez_2015} assumes that the model parameters including the noise precision in the signal model, 
the hyperparameters in the prior distribution on the sparse vector, and the hyperparameters in the prior distribution on 
the support vector are known a priori. For practical massive MIMO channel, 
these parameters are unknown and need to be estimated. One way to estimate the model parameters is by maximizing the marginal likelihood function$-$the procedure which 
is known as type-II maximum likelihood method or evidence 
procedure \cite{Tipping_2001}. However, directly maximizing the marginal likelihood function does not result in closed-form 
update equations for the model parameters \cite{SBL_1,Tipping_2001}. Thus we derive an expectation maximization (EM) 
algorithm which results in closed-form update equations and iteratively computes the maximum likelihood solution of the model parameters \cite{Dempster_EM, Bishop_2006}. 

{\item In order to integrate the EP algorithm with the EM algorithm, we use a variational EM approach \cite{VEM_2008, Murphy_2012} 
in which the approximated joint posterior distribution by the EP algorithm is used to compute the expectation step 
in the EM algorithm. 
The convergence of the resulting EM-EP algorithm is guaranteed through the convergence properties of the variational EM algorithm \cite{VEM_2008}. As iterations 
of the proposed method proceed, the EM algorithm converges to a 
local maxima of the marginal likelihood function \cite{Dempster_EM} and the EP algorithm 
closely approximates the true joint posterior distribution with a 
distribution from an exponential family \cite{MinkaPHD}. 
Grid refining procedure from \cite{nU_Burst_Dai_2019, off_grid1} is also integrated in the proposed EM-EP algorithm to reduce the channel modeling error.} 

{\item Extensive simulations are carried out to demonstrate the efficacy 
of the proposed EM-EP algorithm.
The results are also compared with those in the literature showing the advantages of the proposed method.}
\end{itemize}

This 
paper is organized as follow. 
Section \ref{sys-model} describes the system model for the FDD-based downlink channel estimation in multi-user massive MIMO 
system. Expectation propagation algorithm for this system is proposed in Section \ref{EP_part}. An expectation maximization algorithm to estimate the model parameters 
and to refine the grid is derived in 
Section \ref{EM_part}.
Simulation results are discussed in 
Section \ref{sim_results}, and Section \ref{conclude_sectn} concludes the paper.

\textit{\textbf{Notations:}} Throughout this paper, small letters $(x)$ are used for 
scalars, bold small letters $(\bdx)$ for vectors, and bold capital letters 
$(\bdX)$ for matrices. $\R$ and $\C$ represent the set of real and complex 
numbers, respectively. 
The superscripts $(.)^T$, $(.)^H$, $(.)^*$, and $(.)^{-1}$ represent transpose, 
Hermitian transpose, complex conjugate, and inverse operations, respectively. 
$\CN(\bdx;\bdmu,\bdSigma)$ denotes complex Gaussian distribution on $\bdx$ with 
mean $\bdmu$ and covariance matrix $\bdSigma$.
$\Bern(x;p)$ denotes a Bernoulli distribution on $x$ with mean $p$. For a complex variable $x$, $\lvert x\rvert$, $\Re\{x\}$ and $\Im\{x\}$ represent its 
modulus, real part and imaginary part, respectively. 
For a probability density function (pdf) $p(.)$, 
$\Exp_{p}$ denotes the expectation operator with respect to $p(.)$. $\delta(x)$ 
is the Kronecker delta function which is equal to $1$ when $x=0$ and is zero 
otherwise. 
$\bdI_N$ denotes the $N \times N$ identity matrix.
Finally, tr$(\bdX)$ and $||\bdx||$ denote the trace of a matrix $\bdX$ and the
 $\l_2$-norm of the vector 
$\bdx$, respectively.

\section{System Model} \label{sys-model}

Consider a single cell massive MIMO system where a BS equipped 
with $G$ antennas serves $K$ users each one having a single antenna. It is assumed
that FDD is used and to
enable the estimation of the DL channels,
the BS broadcasts a sequence 
of $N$ pilot symbols denoted by $\bdX=[\bdx_1,\bdx_2,\ldots,\bdx_N]^H$ where 
$\bdx_n \in \C^{G\times 1}$ for $n=1,\ldots,N$. The signal received by the $k$-th 
user is given by 
\begin{equation}\label{kth_recv_sig}
\bdy_k=\bdX \bdh_k+\bdn_k,
\end{equation}
where $\bdy_k\in \C^{N\times 1}$,
$\bdh_k \in \C^{G \times 1}$ is the DL channel to the $k$-th user 
and the receiver noise $\bdn_k$ is distributed as 
$\CN(\bdn_k;\bdzero,\eta^{-1}_k\bdI_N)$ in which 
$\eta_k$ denotes the precision. 

Assuming that the transmitted pilot sequence 
satisfies $\mathrm{tr}(\bdX\bdX^H)=NG$,
the signal-to-noise ratio (SNR) is given by 
$SNR=\eta_k$. 
Suppose that the BS is equipped with a uniform linear array 
(ULA)\footnote{In this work we assume a 
ULA at the BS. However, the 
proposed algorithm can be extended to an arbitrary $2$-D array using the 
approach suggested in \cite{off_grid1}.} 
and to transmit in the direction $\theta$, it uses the beam steering 
vector
\begin{equation}\label{beam_steer}
\bda(\theta)=\left[1, e^{-j2\pi \frac{d}{\lambda_d} \text{sin}(\theta)}, \ldots,e^{-j2\pi \frac{d}{\lambda_d}(G-1) \text{sin}(\theta)}\right]^T,
\end{equation}
where $d$ is the spacing between adjacent antenna elements 
and $\lambda_d$ is the wavelength of the DL signal. 
Let the DL signal propagating from BS on the way 
to the $k$-th user pass across a total of $L_s$ 
scatterers each one forwarding the signal on $L_p$ paths towards the user. 
Then the channel vector 
$\bdh_k$ to the $k$-th user can be written as 
\begin{equation} \label{physical_channel}
\bdh_k=\sum^{L_s}_{s=1}\sum^{L_p}_{p=1}\alpha_{k,s,p}\bda(\theta_{k,s,p}),
\end{equation}
where $\alpha_{k,s,p}$ is the complex path gain for the $s$-th scatterer and 
$p$-th path, and $\theta_{k,s,p}$ is the corresponding AoD \cite{Tse_2005,Bajwa_CS}. 

To reduce the pilot-overhead for estimating this downlink channel, we use the CS approach which requires a virtual channel representation of the physical 
channel in \eqref{physical_channel}. To this end, 
let $\bdtheta=\left(\theta_1,\theta_2,\ldots,\theta_M\right)^T$ 
denote a uniform sampling of the 
interval $[-\pi/2, \pi/2]$ into $M$ points. 
Assuming $M$ is large enough such that the physical AoDs in \eqref{physical_channel} lie on the grid points, the virtual representation of $\bdh_k$ is given by 
\begin{equation}\label{virtual_chan}
\bdh_k=\bdA(\bdtheta) \bdw_k,
\end{equation}
where $\bdA(\bdtheta)=\left[\bda \left({\theta}_{1}\right),\bda \left({\theta}_{2}\right),\ldots,\bda \left({\theta}_{M}\right)\right]$ and 
the vector $\bdw_k$ contains the channel coefficients
in the virtual angular domain. Note that when $M=G$ and the grid is uniformly sampled, the dictionary 
$\bdA(\bdtheta)$ 
represents the unitary discrete 
Fourier transform matrix \cite{off_grid1}. The choice of the parameter $M$ is discussed in Section \ref{sim_results}.

In this paper, we focus on the DL channel estimation for a reference user. Therefore 
dropping the index $k$, from \eqref{kth_recv_sig} 
and \eqref{virtual_chan}, the received signal is written as
\begin{equation}\label{recv_sig}
\bdy=\bdPhi(\bdtheta) \bdw + \bdn,
\end{equation}
in which $\bdPhi(\bdtheta)=\bdX \bdA(\bdtheta)$. 
From \eqref{recv_sig} the likelihood function of $\bdw$ is given  
as $p(\bdy|\bdPhi(\bdtheta),\bdw,\eta)$$=\CN(\bdy;\bdPhi(\bdtheta) \bdw,\eta^{-1}\bdI_N)$.
Given $\bdy$ and $\bdPhi(\bdtheta)$ we aim to compute the posterior distribution of the sparse vector $\bdw$. 
Note that the posterior distribution on $\bdw$ can be used to find the minimum mean squared error (MMSE) estimate of $\bdw$ from which the physical channel estimate is obtained using
\eqref{virtual_chan}.

According to the GSCM model \cite{GSCM_2003}, there are only a few dominant scatterers in the channel, i.e., $L_s$ is small. Moreover,
the forwarding paths from each scatterer are concentrated in a 
small angular spread around the line of sight direction between the BS and the scatterer 
\cite{Overcomp_DFT1,Training_2016}. Thus, $\bdw$ exhibits a clustered sparse structure 
with unknown marking of cluster boundaries.
Hence the support (indices of non-zero elements) of $\bdw$ is unknown 
\cite{off_grid1, Super_Markov_1, nU_Burst_Dai_2019}.
To model the clustered sparse structure of $\bdw$ and to determine 
its support, we condition the $m$-th element of $\bdw$ on a latent 
variable $z_m \in \{0,1\}$, where $w_m \neq 0$ when $z_m=1$ and $w_m=0$ 
when $z_m=0$. Thus given the latent vector 
$\bdz=[z_1,z_2,\ldots,z_M]^T$, as in \cite{S_TCS_paper, Super_Markov_1, Hernandez_2013, Hernandez_2015},
the prior distribution on $\bdw$ is written as 

\begin{align}\label{pw_giv_z}
p(\bdw  |\bdz,\bdgamma)=\prod^{M}_{m=1} p(w_m|z_m,\gamma_m)
=\prod^M_{m=1}\left[z_m \CN(w_m; 0,\gamma^{-1}_m)+(1-z_m)\delta(w_m)\right],
\end{align}
where $\bdgamma=(\gamma_1,\gamma_2,\ldots,\gamma_M)^T$ and $\gamma_m$ 
is the precision of $w_m$. 
Due to the clustered sparsity of $\bdw$, the elements of the
vector $\bdz$ are correlated. To capture this correlation we model $\bdz$ as a 
first-order Markov process with transition probabilities 
$Pr(z_m=1|z_{m-1}=0)=\tau_{01}$ and $Pr(z_m=0|z_{m-1}=1)=\tau_{10}$. Note that these 
transition probabilities reflect the clustered sparse structure of $\bdw$ in the following way. The average length of the sequence of zeros between two consecutive non-zero clusters is large when $\tau_{01}$ is small, and the non-zero cluster size on average is large when $\tau_{10}$ is small. Denoting $\btau \triangleq (\tau_{01}, \tau_{10})$,
the prior 
distribution on $\bdz$ is given as
\begin{align}\label{pr_z}
&p(\bdz|\btau)=p(z_1)\prod^{M}_{m=2} p(z_m|z_{m-1},\btau)\nonumber\\
&=p(z_1)\prod^{M}_{m=2}\left[\left((1-\tau_{10})^{z_{m-1}}(\tau_{01})^{(1-z_{m-1})}\right)^{z_m}\right.
\left.\left((\tau_{10})^{z_{m-1}}(1-\tau_{01})^{(1-z_{m-1})}\right)^{(1-z_m)}\right],
\end{align}
where $p(z_1)=\Bern(z_1;\lambda)$ and we use the steady state distribution for $z_1$ and set 
$\lambda=\frac{\tau_{01}}{\tau_{01}+\tau_{10}}$. 
\begin{figure*}[ht]
    \centering
     	\includegraphics[width=0.99\textwidth]{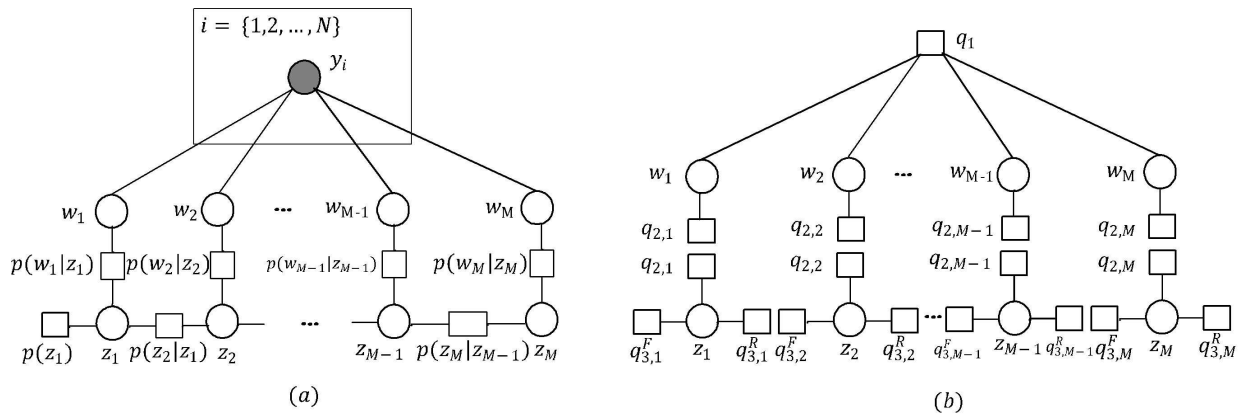}
	\caption{Factor graph illustrations of $(a)$ True posterior distribution in \eqref{opt1b}, and 
	$(b)$ Approximated posterior distribution in \eqref{Qwz}. Variable 
	nodes are represented by circles (filled in circles for observed variables and
	 empty ones for the hidden variables) 
	and factor nodes are denoted by small rectangles. Repetition of observed 
	variables in the subgraph is represented using a plate (big rectangle) notation.}
	\label{fig:True_and_Approx_Posterior}
\end{figure*}

In practice the physical AoDs may not lie on the assumed angular grid $\bdtheta$ in \eqref{virtual_chan}, and thus we 
treat $\bdtheta$ as an unknown parameter and aim to estimate it for learning the dictionary. Therefore, 
letting $\bdxi\triangleq (\btau,\gamma_1,\gamma_2,
\ldots,\gamma_M,\eta,\bdtheta^T)^T$, we aim to jointly estimate $(\bdw,\bdz, \bdxi)$. 
We write the joint posterior distribution of $(\bdw,\bdz, \bdxi)$ as
\begin{align}
p(\bdw,\bdz,\bdxi|\bdy)
& \propto p(\bdw,\bdz|\bdy,\bdxi)p(\bdy|\bdxi)p(\bdxi),\label{post1}
\end{align}
where
the conditioning on $\bdxi$ in \eqref{post1} 
removes the multidimensional integration 
over $\bdxi$ required otherwise 
in computing 
the normalization constant. Note that in \eqref{post1}, the marginal joint posterior distribution on $\bdw$ and $\bdz$ is given by 
\begin{align}\label{opt1b}
p(\bdw,\bdz|\bdy,{\bdxi}) \propto p(\bdy|\bdPhi(\bdtheta),\bdw,{\eta})p(\bdw|\bdz,{\bdgamma})p(\bdz|\btau),
\end{align} 

Computing the joint posterior distribution in \eqref{post1} is still involved. 
We can reduce \eqref{post1} to \eqref{opt1b} by using 
the maximum a posteriori estimate of ${\bdxi}$ in \eqref{opt1b} obtained 
by maximizing $p(\bdy|\bdxi)p(\bdxi)$ with respect to $\bdxi$. 
Assuming a uniform prior distribution on $\bdxi$, we get
the maximum likelihood (ML) estimate which 
can be computed as follows.
\begin{align}\label{opt2}
\quad
\hat{\bdxi}
&=\argmax_{\bdxi} p(\bdy|\bdxi),
\end{align}
The objective function in \eqref{opt2} is a non-concave 
function and due to the involved multidimensional parameter space a brute-force search is difficult \cite{Tipping_2001}. An alternative is to use the iterative expectation maximization 
(EM) algorithm which increases the likelihood function $p(\bdy|\bdxi)$ in each iteration and guarantees convergence to a local maxima \cite{Bishop_2006, Dempster_EM}. 
To this end we define the complete data as $\bdd=[\bdy^T,\bdw^T,\bdz^T]^T$. 
Then if $\bdxi^l$ is the estimate from the $l$-th iteration, in the 
$(l+1)$-st iteration of EM we perform the following two steps
\begin{align}\label{Estep}
&\text{E-Step}: \ \bdL(\bdxi;\bdxi^l)=\Exp_{p(\bdw,\bdz|\bdy,\bdxi^l)}\left[\ln p(\bdy,\bdw,\bdz|\bdxi)\right],\\
&\text{M-Step}: \  \bdxi^{(l+1)}=\argmax_{\bdxi} \bdL(\bdxi;\bdxi^l),\label{Mstep}
\end{align}
and \eqref{Estep} and \eqref{Mstep} are repeated until convergence.

Computing the E-step in \eqref{Estep} requires the exact joint posterior 
distribution $p(\bdw,\bdz|\bdy,\bdxi^l)$ 
which is computationally intractable as it requires  
a multidimensional integration and summation. 
Therefore in Section \ref{EP_part} we derive 
an expectation propagation (EP) algorithm to approximate this 
distribution with a distribution from an exponential family. 
We denote the approximate distribution by $Q(\bdw,\bdz|\bdy,\bdxi^l)$ and use it 
in place of $p(\bdw,\bdz|\bdy,\bdxi^l)$ in \eqref{Estep}.
Note that the estimate of the parameters in the $l$-th iteration of EM, namely $\bdxi^l$
is used by the EP algorithm to obtain $Q(\bdw,\bdz|\bdy,\bdxi^l)$.
Once the E-step is solved in this way, the solution to the M-step,
derived in Section \ref{EM_part}, is computed to obtain $\bdxi^{l+1}$.
Next the EP algorithm is run with $\bdxi^{l+1}$ to obtain 
$Q(\bdw,\bdz|\bdy,\bdxi^{l+1})$ which is used in the $(l+1)$-st iteration of E-step.
The iterations between EM and EP are continued in this way until convergence is achieved. 
This EM-EP approach is reminiscent of the variational EM algorithm 
\cite{VEM_2008, Murphy_2012}. 
We should point out that convergence of EM-EP is assured based on the convergence properties of variational EM \cite{VEM_2008}. 
In particular, as the iterations of the EM-EP proceed,
the EM algorithm converges to a 
local maxima of the objective function in \eqref{opt2} \cite{Dempster_EM} 
and the EP algorithm closely approximates the true joint posterior distribution $p(\bdw,\bdz|\bdy,\bdxi^l)$ in \eqref{opt1b} \cite{MinkaPHD}. 
An EM-EP algorithm has been used in \cite{EM-EP_2006} to solve a classification problem, whereas here we tend to 
use the setting 
for solving the estimation problem.

\section{Expectation Propagation algorithm}\label{EP_part}

In this section, we derive an expectation propagation algorithm to approximate the joint 
posterior distribution $p(\bdw,\bdz|\bdy,\bdxi)$ in \eqref{opt1b} with a distribution from an exponential family. 
For a review of the EP 
algorithm we refer the reader to \cite{Seeger_2005, Qi_2007, CS_EP, Hernandez_2013, Hernandez_2015}. 

Let $\famF$ denote the family of exponential distributions. Exploiting the 
factorized structure of \eqref{opt1b}, we approximate the joint posterior distribution 
$p(\bdw,\bdz|\bdy,\bdxi)$ with 
\begin{equation}\label{EP_approx}
Q(\bdw,\bdz)=Q(\bdw)Q(\bdz),
\end{equation}
where $Q(\bdw)\in \famF$ and $Q(\bdz)\in \famF$\footnote{The conditioning on $\bdy$ and $\bdxi$ is dropped in this section occasionally for 
notational convenience}. We choose the factors in \eqref{EP_approx} as 
\begin{align}\label{Qw}
Q(\bdw)&=\CN(\bdw; \bdmu,\bdSigma),\\
Q(\bdz)&=\prod^M_{m=1}Q_m(z_m)=\prod^M_{m=1}\Bern(z_m;\sigma(p_m)),\label{Qz}
\end{align}
where the sigmoid function $\sigma(.)$ is used to define the mean of 
the Bernoulli distribution as $\sigma(p_m)$\footnote{For a variable $x\in \R$, the sigmoid 
function is defined as $\sigma(x)=\frac{1}{1+e^{-x}}$.}. The use of sigmoid function 
simplifies EP updates and avoids numerical underflow errors resulting in the numerical stability of EP algorithm \cite{Hernandez_2013}. In \eqref{Qw} and 
\eqref{Qz}, $\bdmu$, $\bdSigma$, and $\bdp\triangleq[p_1,p_2,\ldots,p_M]^T$ are 
the unknown parameters that we next aim to estimate with the EP algorithm.

Next we approximate each factor in \eqref{opt1b}. Let
$q_1(\bdw)$, $q_2(\bdw,\bdz)$ and $q_3(\bdz)$
approximate \\
$p(\bdy|\bdPhi(\bdtheta),\bdw,\eta)$,
$p(\bdw|\bdz,\bdgamma)$ and $p(\bdz|\btau)$, respectively. 
Since $q_1(.)$ and $q_3(.)$ are the marginal functions of $\bdw$ and 
$\bdz$, respectively, whereas $q_2(.)$ is the joint function of both $\bdw$ and $\bdz$, we choose these terms as follows


\begin{align}\label{q1w}
q_1(\bdw)&=\CN(\bdw;\bdmu_1,\bdSigma_1),
\end{align}
\begin{align}\label{q2wa}
q_2(\bdw,\bdz)&=\prod^M_{m=1}q_{2,m}(w_m,z_m),
\end{align}
where 
\begin{align}\label{q2wb}
q_{2,m}(w_m,z_m)\propto \CN(w_m;\mu_{2,m},\Sigma_{2,m})\Bern(z_m;\sigma(p_{2,m})),
\end{align}
For $q_3(\bdz)$, we approximate $p(z_m|z_{m-1})$ in \eqref{pr_z} 
with $q^{FR}_{3,m-1,m}(z_{m-1},z_m)$ which in factorized form we write as
$q^{FR}_{3,m-1,m}(z_{m-1},z_m)= q^R_{3,m-1}(z_{m-1}) q^F_{3,m}(z_m)$. 
Therefore
\begin{align}
q_3(\bdz)
&=\prod^M_{m=1}q^R_{3,m}(z_m)q^F_{3,m}(z_m),\label{q3FR}
\end{align}
where for $j\in\{F,R\}$, 
$q^j_{3,m}(z_m)$$=\Bern\big(z_m;\sigma(p^j_{3,m})\big)$ and $\sigma(p^j_{3,m})$ denotes the mean of the Bernoulli distribution. These means actually define the forward and reverse messages sent between $z_{m-1}$ and $z_m$ in the factor graph of 
Fig. \ref{fig:True_and_Approx_Posterior}$(a)$ to get the 
approximate posterior distribution in Fig. \ref{fig:True_and_Approx_Posterior}$(b)$. 
Note that in
\eqref{q3FR} we use the convention that $q^F_{3,1}(z_1)=p(z_1)$ and $q^R_{3,M}(z_M)=1$. 
Next to find the unknown parameters in \eqref{Qw} and \eqref{Qz}, we write 
\begin{equation}\label{Qwz}
Q(\bdw,\bdz)\propto q_1(\bdw) q_2(\bdw,\bdz) q_3(\bdz),
\end{equation}
and using \eqref{q1w}-\eqref{q3FR} in 
\eqref{Qwz} above, we get
\begin{align}\label{bdV}
\bdSigma&=\left(\bdSigma^{-1}_1+\bdSigma^{-1}_2\right)^{-1},\\
\bdmu&=\bdSigma\left(\bdSigma^{-1}_1\bdmu_1+\bdSigma^{-1}_2\bdmu_2\right),\label{bdm}\\
p_m&=\left\{
\begin{matrix}
p_{2,m}+p^F_{3,m}+p^R_{3,m}, &\text{for }m=1,2,\ldots,M-1,\\
p_{2,m}+p^F_{3,m}, &\text{for }m=M,
\end{matrix}\right.\label{pm}
\end{align}
where $\bdmu_2$$=(\mu_{2,1},\mu_{2,2},\ldots,\mu_{2,M})^T$ and $\bdSigma_2$ is a diagonal matrix with $m$-th entry as $[\bdSigma_2]_{m,m}=\Sigma_{2,m}$. 
Note that $p^j_{3,m}$ for $j\in \{F,R\}$ and $p_m$, $p_{2,m}$ in \eqref{pm} are 
the arguments to the sigmoid functions and not the success probabilities of the Bernoulli distributions. Thus, the value of $p_m$ in \eqref{pm} can be outside the range $[0,1]$.
However, the output of the sigmoid
function with input $p_m$ will be in the range $[0, 1]$ representing
the success probability{\footnote{
To derive \eqref{pm}, we used the following facts.
Firstly, $\prod^N_{n=1}\Bern(x;\phi_n) \propto \Bern(x;\phi)$ where 
$\phi=\frac{\prod^N_{n=1}\phi_n}{\prod^N_{n=1}\phi_n+\prod^N_{n=1}(1-\phi_n)}$. 
Secondly, the inverse sigmoid (logit) function is given by 
$\sigma^{-1}(x)=\ln \frac{x}{1-x}$.}.
Also note that since in \eqref{q3FR} we set $q^F_{3,1}(z_1)=p(z_1)$, this implies 
that in \eqref{pm} $p^F_{3,1}=\sigma^{-1}(\lambda)$. 
Both the true posterior distribution in \eqref{opt1b} and the approximated one 
in \eqref{Qwz} are depicted in Fig. \ref{fig:True_and_Approx_Posterior} for clarity.
\begin{figure}[tp]
        \centering
     	\includegraphics[width=0.36\textwidth]{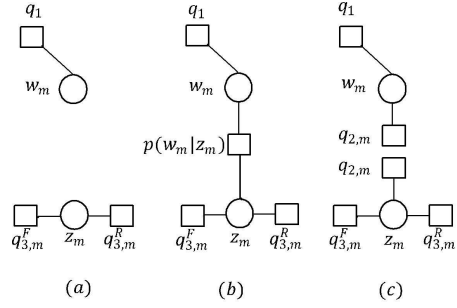}
	\caption{EP steps for updating $q_{2,m}(w_m,z_m)$: $(a)$ Eliminate $q_{2,m}(w_m,z_m)$ from the factor graph 
to find the cavity distribution $Q_{\backslash 2m}(w_m,z_m)$ as in \eqref{Qb2m}, $(b)$ Use $p(w_m|z_m)$ factor to define the 
hybrid posterior distribution $R_{2,m}(w_m,z_m)$ as in \eqref{hybrid_2m}, and 
$(c)$ Project $R_{2,m}(w_m,z_m)$ onto $\famF$ and update 
$q_{2,m}(w_m,z_m)$ as in \eqref{KL2_1}, \eqref{updated_q2w}, and \eqref{updated_q2z}.}
	\label{fig:Q2factor_computation}
\end{figure}
\begin{figure}[tp]
        \centering
     	\includegraphics[width=0.38\textwidth]{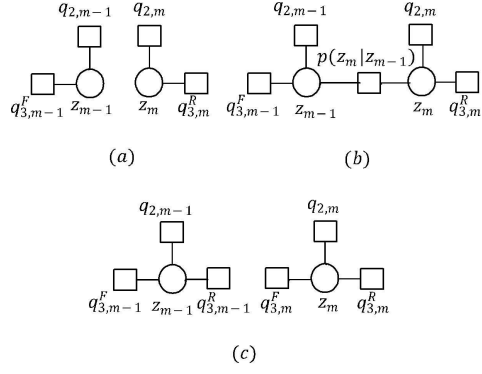}
	\caption{EP steps for updating $q^R_{3,m-1}(z_{m-1})$ and $q^F_{3,m}(z_m)$: $(a)$ Eliminate $q^R_{3,m-1}(z_{m-1})$ and $q^F_{3,m}(z_m)$ from the factor graph to find the cavity distributions $q^{\backslash R}_{3,m-1}(z_{m-1})$ and $q^{\backslash F}_{3,m}(z_m)$ as in \eqref{qbR3z} and \eqref{qbF3z}, $(b)$ Use $p(z_m|z_{m-1})$ factor to define the hybrid posterior distribution $S_{3,m-1,m}(z_{m-1},z_m)$ as in \eqref{hybrid_zm}, and $(c)$ Project $S_{3,m-1,m}(z_{m-1},z_m)$ onto $\famF$ and update $q^R_{3,m-1}(z_{m-1})$ and $q^F_{3,m}(z_m)$ as in \eqref{KL3_a}-\eqref{updated_F3mz}.}
	\label{fig:Forward_Reverse_computation}
\end{figure}

Now as $q_1(\bdw)$ approximates 
$p(\bdy|\bdPhi(\bdtheta),\bdw,\eta)$ which is a complex Gaussian function of $\bdw$ then to simplify we  
set $q_1(\bdw)\propto \CN(\bdy;\bdPhi\bdw,\eta^{-1}\bdI_N)$. Expanding this Gaussian 
distribution and completing the square for $\bdw$, we get 
\begin{align}\label{q1_approx}
\bdSigma^{-1}_1=\eta\bdPhi^H \bdPhi,\quad \bdSigma^{-1}_1\bdmu_1=\eta\bdPhi^H\bdy,
\end{align}
using \eqref{q1_approx}, \eqref{bdV} and \eqref{bdm} can be simplified as 
\begin{align}\label{Eq_bdV}
\bdSigma&=\bdSigma_2-\bdSigma_2\bdPhi^H\left(\eta^{-1}\bdI_N+\bdPhi\bdSigma_2\bdPhi^H\right)^{-1}
\bdPhi\bdSigma_2,\\
\bdmu&=\bdSigma\left(\eta\bdPhi^H\bdy+\bdSigma^{-1}_2\bdmu_2\right),
\label{Eq_bdm}
\end{align} 
Thus to compute \eqref{pm}, \eqref{Eq_bdV}, and \eqref{Eq_bdm} we just need to  update the approximation factors $q_2(\bdw,\bdz)$ 
and $q_3(\bdz)$. We first update $q_2(\bdw,\bdz)$ as follow. Since it is equal to the product 
of marginals $q_{2,m}(w_m,z_m)$, we can instead update each marginal distribution 
individually and in parallel \cite{Seeger_2005}. The steps involved in upating $q_{2,m}(w_m,z_m)$ are depicted in Fig. \ref{fig:Q2factor_computation}}. 
Let $Q_m(w_m,z_m)$ denote the 
marginal distribution obtained from \eqref{EP_approx}. 
Then using \eqref{Qw} and \eqref{Qz} we can write
\begin{align}\label{Q_m_decomposed}
Q_m(w_m,z_m)&=Q_m(w_m)Q_m(z_m),\nonumber\\
&\propto \CN(w_m;\mu_m,\Sigma_{m,m})\Bern(z_m;\sigma(p_m)),
\end{align}
where $\mu_m$ is the $m$-th element of $\bdmu$, and 
$\Sigma_{m,m}=[\bdSigma]_{m,m}$, $m=1,2,\ldots,M$. Following 
the EP framework we first find the cavity distribution as
\begin{align}\label{Qb2m}
Q_{\backslash 2,m}(w_m,z_m)&=\frac{Q_m(w_m,z_m)}{q_{2,m}(w_m,z_m)}
\propto Q_{\backslash 2,m}(w_m)Q_{\backslash 2,m}(z_m),
\end{align}
where $Q_{\backslash 2,m}(w_m)=\CN(w_m;\mu_{\backslash 2,m},\Sigma_{\backslash 2,m})$ and $Q_{\backslash 2,m}(z_m)=\Bern(z_m;\sigma(p_{\backslash 2,m}))$. 
The parameters in these distributions are given by\footnote{To derive \eqref{pb2m} 
we use the fact that for a Bernoulli variable $x$, we have $\frac{\Bern(x;\phi_1)}{\Bern(x;\phi_2)} \propto \Bern(x;\phi)$ where $\phi=\frac{\phi_1/\phi_2}{\phi_1/\phi_2+(1-\phi_1)/(1-\phi_2)}$.}
\begin{align} 
\Sigma_{\backslash 2,m}&=\left(\Sigma^{-1}_{m,m}-\Sigma^{-1}_{2,m}\right)^{-1},\label{vb2m}\\
\mu_{\backslash 2,m}&=\Sigma_{\backslash 2,m}\left(\Sigma^{-1}_{m,m}\mu_m-
\Sigma^{-1}_{2,m}\mu_{2,m}\right),\label{mb2m}\\
p_{\backslash 2,m}&=p_m-p_{2,m},\label{pb2m}
\end{align}

Next we define the hybrid posterior distribution $R_{2,m}(w_m,z_m)$ as
\begin{align}\label{hybrid_2m}
R_{2,m}(w_m,z_m)&=\frac{1}{C_m}p(w_m|z_m)Q_{\backslash 2,m}(w_m,z_m),
\end{align}
where $p(w_m|z_m)$ is defined in \eqref{pw_giv_z}. The normalization constant $C_m$ 
in \eqref{hybrid_2m} is computed as follow
\begin{align}
C_m&=\sum_{z_m\in\{0,1\}}\int p(w_m|z_m)Q_{\backslash 2,m}(w_m,z_m)d{w_m},\nonumber\\
&=\int \CN(w_m;0,\gamma^{-1}_m)\CN(w_m;\mu_{\backslash 2,m},\Sigma_{\backslash 2,m})dw_m\sigma(p_{\backslash 2,m}) \nonumber\\
&\quad +\int \delta(w_m)\CN(w_m;\mu_{\backslash 2,m},\Sigma_{\backslash 2,m})dw_m
(1-\sigma(p_{\backslash 2,m})),\nonumber\\
&=\CN(0;\mu_{\backslash 2,m},\Sigma_{\backslash 2,m}+\gamma^{-1}_m)\sigma(p_{\backslash 2,m})+
\CN(0;\mu_{\backslash 2,m},\Sigma_{\backslash 2,m})(1-\sigma(p_{\backslash 2,m})),\label{norm_const}
\end{align}

We now update the approximation $Q_m(w_m,z_m)$ by projecting  
$R_{2,m}(w_m,z_m)$ onto the closest distribution in $\famF$ by minimizing 
the following Kullback-Leibler (KL) divergence
\begin{align}\label{KL2_1}
Q_m(w_m,z_m)
=\argmin_{Q_m(w_m,z_m)\in \famF} KL(R_{2,m}(w_m,z_m)\|Q_m(w_m,z_m)),
\end{align}
since $Q_m(w_m,z_m)=Q_m(w_m)Q_m(z_m)$ from \eqref{Q_m_decomposed}, 
it can be shown that the   
optimization problem in \eqref{KL2_1} is equivalent to solving the following two separate problems \cite{Ghavami3}
\begin{align}\label{KL2_1a}
Q_m(w_m)
&=\argmin_{Q_m(w_m)\in \famF} KL\left(R_{2,m}(w_m)\|Q_m(w_m)\right),
\end{align}
and
\begin{align}\label{KL2_1b}
Q_m(z_m)
&=\argmin_{Q_m(z_m)\in \famF} KL\left(R_{2,m}(z_m)\|Q_m(z_m)\right),
\end{align}
where $R_{2,m}(w_m)$ and $R_{2,m}(z_m)$ are the marginal distributions of $R_{2,m}(z_m)$. 
The KL divergence in \eqref{KL2_1a} and \eqref{KL2_1b} is minimized by 
using the moment matching property \cite{Hernandez_2015}. Thus  
for $Q_m(w_m)$ and $Q_m(z_m)$ defined in \eqref{Q_m_decomposed} we set 
\begin{align}\label{meanRw}
\mu_m&=\Exp_{R_{2,m}}[w_m],\\
\Sigma_{m,m}&=\Exp_{R_{2,m}}[\lvert w_m\rvert^2]-
\lvert\Exp_{R_{2,m}}[w_m]\rvert^2,\label{varRw}\\
\sigma(p_m)&=\Exp_{R_{2,m}}[z_m],\label{meanRz}
\end{align}
The values of $\mu_m$, $\Sigma_{m,m}$, and $\sigma(p_m)$ are given in the 
following lemma which is proved in Appendix A.
\begin{lemma}\label{lemma1}
\begin{enumerate}
\item
The posterior mean value $\sigma(p_m)$ is given by 
\begin{align}\label{lem1_sig_Eq}
\sigma(p_m)
=\left(1+\frac{\sigma(-p_{\backslash 2,m})\CN(0;\mu_{\backslash 2,m},\Sigma_{\backslash 2,m})}{\sigma(p_{\backslash 2,m})\CN(0;\mu_{\backslash 2,m},\Sigma_{\backslash 2,m}+\gamma^{-1}_m)}\right)^{-1},
\end{align}

\item
The posterior mean value $\mu_m$ is given by
\begin{align}\label{lem1_mu}
\mu_m=\mu_{\backslash 2,m}+\Sigma_{\backslash 2,m}\frac{\partial \ln C_m}{\partial \mu^*_{\backslash 2,m}},
\end{align}
where
\begin{align}
\frac{\partial \ln C_m}{\partial \mu_{\backslash 2,m}}=-\sigma(p_m)\frac{\mu^*_{\backslash 2,m}}{\Sigma_{\backslash 2,m}+\gamma^{-1}_m}-\sigma(-p_m)\frac{\mu^*_{\backslash 2,m}}{\Sigma_{\backslash 2,m}},
\end{align}
\item
The posterior variance $\Sigma_{m,m}$ is given by
\begin{align}\label{lem1_var}
\Sigma_{m,m}
=\Sigma_{\backslash 2,m}+(\Sigma_{\backslash 2,m})^2
\left[\frac{\partial \ln C_m}{\partial \Sigma_{\backslash 2,m}}-
\frac{\partial \ln C_m}{\partial \mu^*_{\backslash 2,m}}\frac{\partial \ln C_m}{\partial \mu_{\backslash 2,m}}\right],
\end{align}
where 
\begin{align}
\frac{\partial \ln C_m}{\partial \Sigma_{\backslash 2,m}}=\sigma(p_m)\frac{|\mu_{\backslash 2,m}|^2-\left(\Sigma_{\backslash 2,m}
+\gamma^{-1}_m\right)}{\left(\Sigma_{\backslash 2,m}
+\gamma^{-1}_m\right)^2}+
\sigma(-p_m)\frac{|\mu_{\backslash 2,m}|^2-\left(\Sigma_{\backslash 2,m}\right)}{\left(\Sigma_{\backslash 2,m}\right)^2},
\end{align}
\end{enumerate}
\end{lemma}
Next we update the factor $q_{2,m}(w_m,z_m)$. Since $q_{2,m}(w_m,z_m)$$=q_{2,m}(w_m)q_{2,m}(z_m)$ 
we can update the marginals separately. To update $q_{2,m}(w_m)$ we write
\begin{align}\label{updated_q2w}
q_{2,m}(w_m)&=\frac{Q_m(w_m)}{Q_{\backslash 2,m}(w_m)}=\frac{\CN(w_m;\mu_m,\Sigma_{m,m})}{\CN(w_m;\mu_{\backslash 2,m},\Sigma_{\backslash 2,m})},\nonumber\\
& \propto \CN(w_m;\mu_{2,m},\Sigma_{2,m}),
\end{align}
where 
\begin{align}\label{updated_Sig2m}
\Sigma_{2,m}&=\left(\left(\Sigma_{m,m}\right)^{-1}-\left(\Sigma_{\backslash 2,m}\right)^{-1}\right)^{-1},\\
\mu_{2,m}&=\Sigma_{2,m}\left(\left(\Sigma_{m,m}\right)^{-1}\mu_m-\left(\Sigma_{\backslash 2,m}\right)^{-1}\mu_{\backslash 2,m}\right),\label{updated_mu2m}
\end{align}
and to update $q_{2,m}(z_m)$ we write 
\begin{align}\label{updated_q2z}
q_{2,m}(z_m)&= \frac{Q_m(z_m)}{Q_{\backslash 2,m}(z_m)}
=\frac{\Bern(z_m;\sigma(p_m))}{\Bern(z_m;\sigma(p_{\backslash 2,m}))}\propto
\Bern(z_m;\sigma(p_{2,m})),
\end{align}
where 
\begin{align}\label{updated_z_2m}
\sigma(p_{2,m})
=\frac{\CN(0;\mu_{\backslash 2,m},\Sigma_{\backslash 2,m}+\gamma^{-1}_m)}{\CN(0;\mu_{\backslash 2,m},\Sigma_{\backslash 2,m}+\gamma^{-1}_m)
+\CN(0;\mu_{\backslash 2,m},\Sigma_{\backslash 2,m})},
\end{align}
and using the logit function $\sigma^{-1}(.)$ on \eqref{updated_z_2m} we get
\begin{align}\label{updated_p2m}
p_{2,m}=\ln \CN(0;\mu_{\backslash 2,m},\Sigma_{\backslash 2,m}+\gamma^{-1}_m)-
 \ln \CN(0;\mu_{\backslash 2,m},\Sigma_{\backslash 2,m})
\end{align}

Next we update the approximation factor $q_3(\bdz)$ in \eqref{Qwz}. We start by updating $q^R_{3,m-1}(z_{m-1})$ and $q^F_{3,m}(z_{m})$. The EP steps taken to update these factors are summarized in Fig. \ref{fig:Forward_Reverse_computation}. Given the 
marginal distribution on $z_m$ as 
$Q_{m}(z_m)=q^F_{3,m}(z_m)q_{2,m}(z_m)q^R_{3,m}(z_m)$ which is also easily observable from Fig. \ref{fig:True_and_Approx_Posterior}(b), we first find the cavity 
distribution $q^{\backslash R}_{3,m-1}(z_{m-1})$ as follow
\begin{align}
q^{\backslash R}_{3,m-1}(z_{m-1})&=\frac{Q_{m-1}(z_{m-1})}{q^R_{3,m-1}(z_{m-1})}
=q^F_{3,{m-1}}(z_{m-1})q_{2,{m-1}}(z_{m-1})\nonumber\\
&\propto \Bern\left(z_{m-1};\sigma\left(p^{\backslash R}_{3,m-1}\right)\right),\label{qbR3z}
\end{align}
where
\begin{align}\label{sig_bRm_a}
\sigma\left(p^{\backslash R}_{3,m-1}\right)
=\frac{\sigma\left(p^{F}_{3,m-1}\right)\sigma\left(p_{2,m-1}\right)}{\sigma\left(p^{F}_{3,m-1}\right)\sigma\left(p_{2,m-1}\right)+\sigma\left(-p^{F}_{3,m-1}\right)\sigma\left(-p_{2,m-1}\right)},
\end{align}
Solving \eqref{sig_bRm_a} using the logit function $\sigma^{-1}(.)$ and adjusting the notation to update the $m$-th factor we get
\begin{align}\label{sig_bRm_b}
p^{\backslash R}_{3,m}&=p_{2,m}+p^F_{3,m}, \quad \text{for} \ m=1,2,\ldots,M-1
\end{align}
Similarly, the cavity distribution $q^{\backslash F}_{3,m}(z_{m})$ can also be found by
\begin{align} 
q^{\backslash F}_{3,m}(z_{m})=\frac{Q_{m}(z_{m})}{q^F_{3,m}(z_{m})}
=q^R_{3,m}(z_m)q_{2,m}(z_m)
\propto \Bern\left(z_{m};\sigma\left(p^{\backslash F}_{3,m}\right)\right)\label{qbF3z}.
\end{align}
Following a similar approach as in \eqref{sig_bRm_a} and \eqref{sig_bRm_b} we get 
\begin{align}\label{sig_bFm}
p^{\backslash F}_{3,m}&=\left\{
\begin{matrix}
&p_{2,m}+p^R_{3,m}, && \text{for} \quad m=1,2,\ldots,M-1\\
&p_{2,m}, && \text{for}  \quad m=M
\end{matrix}\right.
\end{align}

Once the cavity distributions are computed, we define the hybrid joint posterior distribution on $z_{m-1}$ and $z_m$ as
\begin{align}\label{hybrid_zm}
S_{3,m-1,m}(z_{m-1},z_m)&=q^{\backslash R}_{3,m-1}(z_{m-1})p(z_m|z_{m-1})
q^{\backslash F}_{3,m}(z_{m}),
\end{align}
in which $p(z_m|z_{m-1})$ is given in \eqref{pr_z}. 
Since \eqref{hybrid_zm} involves a product of Bernoulli distributions, 
$S_{3,m-1,m}(z_{m-1},z_m)$ is a bivariate Bernoulli distribution where 
the marginal distributions on $z_{m-1}$ and $z_m$ can be written as 
\begin{align}\label{S3z_a}
S_{3,m-1}(z_{m-1})&=\sum_{z_m\in \{0,1\}}S_{3,m-1,m}(z_{m-1},z_m),\\ 
S_{3,m}(z_{m})&=\sum_{z_{m-1}\in \{0,1\}}S_{3,m-1,m}(z_{m-1},z_m),\label{S3z_b}
\end{align}
and using their derived forms in Appendix B the means of these marginal Bernoulli distributions are found as
\begin{align}
\Exp_{S_{3,m-1}}[z_{m-1}]
=\frac{1}{D_m}\sigma(p^{\backslash R}_{3,m-1})\left[
\sigma(p^{\backslash F}_{3,m})(1-\tau_{10})+\sigma(-p^{\backslash F}_{3,m})
\tau_{10}\right],\label{mean_S_3a}\\
\Exp_{S_{3,m}}[z_m]
=\frac{1}{D_m}\sigma(p^{\backslash F}_{3,m})\left[
\sigma(p^{\backslash R}_{3,m-1})(1-\tau_{10})+\sigma(-p^{\backslash R}_{3,m-1})
\tau_{01}\right]\label{mean_S_3b},
\end{align}
where the normalization constant $D_m$ is given by
\begin{align}\label{norm_const_Dm}
D_m
&=\sigma(-p^{\backslash R}_{3,m-1})\sigma(-p^{\backslash F}_{3,m})(1-\tau_{01})
+\sigma(p^{\backslash R}_{3,m-1})\sigma(-p^{\backslash F}_{3,m})\tau_{10}
\nonumber\\
& +\sigma(-p^{\backslash R}_{3,m-1})\sigma(p^{\backslash F}_{3,m})\tau_{01}+
\sigma(p^{\backslash R}_{3,m-1})\sigma(p^{\backslash F}_{3,m})(1-\tau_{10}),
\end{align}

Now we update the approximation factors $Q_{m-1}(z_{m-1})$ and $Q_{m}(z_m)$ by projecting \\
$S_{3,m-1,m}(z_{m-1},z_m)$  in \eqref{hybrid_zm} onto the closest distribution in $\famF$. This is done 
by minimizing the KL divergence between $S_{3,m-1,m}(z_{m-1},z_m)$ and $Q_{m-1}(z_{m-1})Q_{m}(z_m)$. As in \eqref{KL2_1}, this can be achieved by solving two 
separate optimization problems  
\begin{align} \label{KL3_a}
Q_{m-1}(z_{m-1})
=\argmin_{Q_{m-1}(z_{m-1})\in \famF} 
KL\left(S_{3,m-1}(z_{m-1})\|Q_{m-1}(z_{m-1})\right),
\end{align}
and
\begin{align}\label{KL3_b}
Q_{m}(z_{m})
&=\argmin_{Q_{m}(z_{m})\in \famF} 
KL\left(S_{3,m}(z_{m})\|Q_{m}(z_{m})\right),
\end{align}
where the marginals $S_{3,m-1}(z_{m-1})$ and $S_{3,m}(z_{m})$ are computed from \eqref{S3z_a} and \eqref{S3z_b}. The KL divergence in \eqref{KL3_a} and 
\eqref{KL3_b} is minimized as before by using the moment matching property. 
Thus we set $\sigma(p_{m-1})=\Exp_{S_{3,m-1}}[z_{m-1}]$ given in \eqref{mean_S_3a} and $\sigma(p_{m})=\Exp_{S_{3,m}}[z_m]$ given in \eqref{mean_S_3b}.

Finally we update the approximation factors $q^R_{3,m-1}(z_{m-1})$ and 
$q^F_{3,m}(z_m)$ as follow. To update $q^R_{3,m-1}(z_{m-1})$ we write 
\begin{align}
q^R_{3,m-1}(z_{m-1})=\frac{Q_{m-1}(z_{m-1})}{q^{\backslash R}_{3,m-1}(z_{m-1})}\propto \Bern(z_{m-1};\sigma(p^R_{3,m-1})),
\end{align}
where $\sigma(p^R_{3,m-1})$ is computed from \eqref{updated_q3R} in which the notation is adjusted to compute the $m$-th factor. Similarly 
to update $q^F_{3,m}(z_m)$ we write
\begin{align}
q^F_{3,m}(z_{m})=\frac{Q_{m}(z_{m})}{q^{\backslash F}_{3,m}(z_{m})}\propto \Bern(z_{m};\sigma(p^F_{3,m})),\label{updated_F3mz}
\end{align}
where $\sigma(p^F_{3,m})$ is computed from \eqref{updated_q3F}. 
This completes all the posterior updates required for an EP's iteration. The complete EP algorithm is summarized in Algorithm \ref{algo1}.
\begin{algorithm}\label{algo1}
  \footnotesize
\DontPrintSemicolon
\SetKwInput{KwPara}{Parameters}
\SetKwFor{ForEach}{for each}{}{end}
\SetKwRepeat{Repeat}{repeat}{until}
\SetKw{Break}{break}
\KwIn{$\bdy$}
\KwPara{$\bdxi$,$\bdtheta$.}
\tcc{EP run}
\ForEach{$n = \{1,2,\ldots,n_{EP}\}$}
{\begin{enumerate}
\item Compute $Q(\bdw,\bdz)$ parameters $\bdp$, $\bdSigma$, and $\bdmu$ using
\eqref{pm}, \eqref{Eq_bdV}, \\ and \eqref{Eq_bdm}, respectively. 
\end{enumerate}
\tcc{Updating factor $q_2(\bdw,\bdz)$:}
\ForEach{$m = \{1,2,\ldots,M\}$}
{
\begin{enumerate}
\item Find $Q_{\backslash 2,m}(w_m,z_m)$ parameters $\Sigma_{\backslash 2,m}$, 
$\mu_{\backslash 2,m}$, and \\
$p_{\backslash 2,m}$ from \eqref{vb2m}, \eqref{mb2m}, and \eqref{pb2m}, respectively. 
\item Update $Q_m(w_m,z_m)$ by computing $p_m$ from \eqref{lem1_sig_Eq},\\ $\mu_m$ from \eqref{lem1_mu}, and $\Sigma_{m,m}$ using \eqref{lem1_var}. 
\item Update the factor $q_{2,m}(w_m, z_m)$ by computing $\Sigma_{2,m}$ \\
from \eqref{updated_Sig2m}, $\mu_{2,m}$ from \eqref{updated_mu2m}, and 
$p_{2,m}$ using \eqref{updated_p2m}.
\end{enumerate}
}
\tcc{Updating factor $q_3(\bdz)$:}
\tcc{Forward pass:}
\ForEach{$m=\{1,2,\ldots,M\}$}
{\begin{enumerate}
\item To update $q^{\backslash R}_{3,m}(z_m)$ factor, compute $p^{\backslash R}_{3,m}$ from \eqref{sig_bRm_b}, \\
if $m<M$.
\item Update $q^F_{3,m}(z_m)$ by computing $p^F_{3,m}$ using \eqref{updated_q3F}, \\ if $m>1$. 
\end{enumerate}
}
\tcc{Reverse pass:}
\ForEach{$m=\{M,M-1,\ldots,1\}$}
{
\begin{enumerate}
\item To update $q^{\backslash F}_{3,m}(z_m)$ factor, compute $p^{\backslash F}_{3,m}$ 
from \eqref{sig_bFm}.
\item Update the factor $q^{R}_{3,m}(z_m)$ by computing $p^R_{3,m}$ from \\ \eqref{updated_q3R}, if 
$m<M$.
\end{enumerate}
}
\tcc{Check for convergence: Keep track of $\bdmu$ for each $n^{th}$ iteration}
\If{$\frac{||\bdmu^n-\bdmu^{n-1}||}{||\bdmu^{n-1}||}<\epsilon_{EP}$}{break;}
}
\KwOut{$\bdmu$, $\bdSigma$, $\bdp$}
\caption{EP Algorithm}
\end{algorithm}
\begin{figure*}[tp]
\normalsize
\setcounter{mytempeqncnt}{\value{equation}}
\setcounter{equation}{\value{mytempeqncnt}}
\begin{align}
\sigma(p^R_{3,m})
&=\frac{\sigma\left(p^{\backslash F}_{3,m+1}\right)(1-\tau_{10})+\sigma\left(-p^{\backslash F}_{3,m+1}\right)
\tau_{10}}
{\sigma\left(p^{\backslash F}_{3,m+1}\right)(1-\tau_{10})+\sigma\left(-p^{\backslash F}_{3,m+1}\right)
\tau_{10}+\sigma\left(p^{\backslash F}_{3,m+1}\right)\tau_{01}+\sigma\left(-p^{\backslash F}_{3,m+1}\right) 
(1-\tau_{01})}, \nonumber \\
&~~~~\text{for } 
~m=1,2,\ldots,M-1,\label{updated_q3R}\\
\sigma(p^F_{3,m})&=\sigma\left(p^{\backslash R}_{3,m-1}\right)(1-\tau_{10})+
\sigma\left(-p^{\backslash R}_{3,m-1}\right)\tau_{01},
 \qquad \text{for } m=2,\ldots,M, \label{updated_q3F}
\end{align}
\hrulefill
\end{figure*}
\begin{remark}
%
In order to improve the convergence of our proposed EP algorithm, when 
$\big(\left(\Sigma_{m,m}\right)^{-1}\\-\left(\Sigma_{\backslash 2,m}\right)^{-1}\big)^{-1}\geq 0$,
we follow the approach suggested in \cite{Hernandez_2015, Seeger_2005} for an EP algorithm, and damp the updates of the factors $\{q_{2,m}(w_m,z_m)\}^M_{m=1}$, $\{q^F_{3,m}(z_m)\}^M_{m=2}$, and $\{q^R_{3,m}(z_m)\}^{M-1}_{m=1}$ in every EP iteration. 
{Using a smoothing mechanism the 
parameters $\Sigma_{2,m}, ~\mu_{2,m}, ~p_{2,m}$ and $ p^{j}_{3,m} $, $j\in\{F,R\}$,
are damped according to the equation
\begin{align}
\psi^{damp} = \beta \psi + (1-\beta) \psi^{old}
\end{align}
where $\beta \in(0,1)$ is the smoothing factor, $\psi^{old}$ represents the parameter in the previous EP iteration and $\psi$ is the value calculated according to the dervations in Section
\ref{EP_part}.} 
The superscript $damp$ denotes the value of the parameter after applying the smoothing mechanism. 
The above damped updates replace the respective undamped ones in the next iteration of EP.  
Further, to improve the convergence of EP 
we use the annealed damping scheme as suggested in \cite{Hernandez_2015} where 
we start the EP algorithm with $\beta=0.5$ and 
progressively anneal its value by multiplying it with a constant $\kappa<1$ after every iteration of EP until convergence. 
Based on empirical evidence we select $\kappa=0.945$ for the considered channel estimation problem in this paper.    
Note that as indicated in \cite{Hernandez_2015} 
we can also have $\left(\left(\Sigma_{m,m}\right)^{-1}-\left(\Sigma_{\backslash 2,m}\right)^{-1}\right)^{-1}<0$ and when this happen we just set $\Sigma_{2,m}=10^2$ and  use the above smoothing mechanism. 
\end{remark}

\section{Expectation Maximization algorithm: E-step and M-step derivations}\label{EM_part}

In this section we evaluate the E-Step and M-step of the EM algorithm as discussed in \eqref{Estep} and \eqref{Mstep}. Using EM 
we aim to iteratively find the ML estimate of the unknown parameters
$\bdxi=(\btau,\gamma_1,\gamma_2,
\ldots,\gamma_M,\eta,\bdtheta)^T$. 
For the complete data defined in section \ref{sys-model} as $\bdd=[\bdy^T,\bdw^T,\bdz^T]^T$ and using the EP's approximation to the posterior distribution 
from \eqref{EP_approx}, the E-step in \eqref{Estep} can be written as
\begin{align}\label{Estep_derive}
&\bdL(\bdxi;\bdxi^l)\approx\Exp_{Q(\bdw,\bdz|\bdy,\bdxi^l)}\left[\ln p(\bdy,\bdw,\bdz|\bdxi)\right],\nonumber\\
&=\Exp_{Q(\bdw,\bdz|\bdy,\bdxi^l)}\left[\ln p(\bdy|\bdPhi(\bdtheta),\bdw,{\eta})p(\bdw|\bdz,{\bdgamma})p(\bdz|{p}_{10},{p}_{01})\right],
\end{align}

Since jointly maximizing \eqref{Estep_derive} over 
$\bdxi$ is difficult, here we instead update $\bdxi$ one element at 
a time while keeping the other elements fixed to their current estimates in the $l$-th iteration 
\cite{Iterative_EM}. To estimate $\tau_{10}$ and $\tau_{01}$, since only 
$p(\bdz|\btau)$ involves these 
parameters, \eqref{Estep_derive} simplifies to
\begin{align}
&\bdL_1(\btau; \btau^l)=\Exp_{Q(\bdw,\bdz|\bdy,\bdxi^l)}\left[\ln p(\bdz|{p}_{10},{p}_{01})\right]\nonumber\\
&=\sum^M_{m=2}\left[\ln (1-\tau_{01})+\sigma\left(p^{(l+1)}_m\right)\sigma\left(p^{(l+1)}_{m-1}\right)\times\right.
\left.\ln \frac{(1-\tau_{10})(1-\tau_{01})}{\tau_{01}\tau_{10}}+\sigma\left(p^{(l+1)}_m\right)\ln \frac{\tau_{01}}{(1-\tau_{01})}\right.\nonumber\\
&\left.+\sigma\left(p^{(l+1)}_{m-1}\right)\ln \frac{\tau_{10}}{(1-\tau_{01})}\right]+\text{const},\label{EM_L1}
\end{align}
where we use the fact that $\Exp_{Q}[z_{m}]=\sigma(p_m)$. 
Maximizing $\bdL_1(.)$ with respect to (w.r.t) $\btau$,
we get the update equations as
\begin{align}\label{updated_p01}
\tau^{(l+1)}_{01}&=\frac{\sum^M_{m=2}\left[\sigma\left(p^{(l+1)}_{m-1}\right)\left(1-\sigma\left(p^{(l+1)}_m\right)\right)\right]}
{\sum^M_{m=2}\sigma\left(p^{(l+1)}_{m-1}\right)},\\
\tau^{(l+1)}_{10}&=\frac{\sum^M_{m=2}\left[\sigma\left(p^{(l+1)}_{m}\right)\left(1-\sigma\left(p^{(l+1)}_{m-1}\right)\right)\right]}
{\sum^M_{m=2}\left(1-\sigma\left(p^{(l+1)}_{m-1}\right)\right)},\label{updated_p10}
\end{align} 

Similarly, maximizing $\bdL(.)$ w.r.t 
$\gamma_m$ and $\eta$ we get 
\begin{align}\label{updated_gamma_m}
\gamma^{(l+1)}_m=\left(\Sigma^{(l+1)}_{m,m}+|\mu^{(l+1)}_m|^2\right)^{-1},
\end{align}
and, 
\begin{align}\label{updated_eta}
\eta^{(l+1)}=\frac{N}{\|\bdy-\bdPhi(\bdtheta^l)\bdmu^{(l+1)}\|^2+\text{tr}\left\{\bdPhi(\bdtheta^l) \bdSigma^{(l+1)} \bdPhi^H
(\bdtheta^l)\right\}},
\end{align}
where to get \eqref{updated_gamma_m} we use the fact that 
$\Exp_{Q}[|w_m|^2]=\Sigma_{m,m}+|\mu_m|^2$ in which $\mu_m$ and $\Sigma_{m,m}$ are defined in \eqref{Q_m_decomposed}, and in \eqref{updated_eta} we use
the fact that  $\Exp_Q[\bdw]=\bdmu$ and $\Exp_Q[\bdw\bdw^H]=\bdSigma+\bdmu\bdmu^H$. Both $\bdSigma$ and $\bdmu$ are given in \eqref{Eq_bdV} and \eqref{Eq_bdm}. 

Finally to update $\bdtheta$ for dictionary learning and minimizing the modeling error, the objective function in \eqref{Estep_derive} can be simplified to
\begin{equation}\label{Obj_L4}
\bdL_2(\bdtheta)=\|\bdy-\bdPhi(\bdtheta)\bdmu^{(l+1)}\|^2+\text{tr}\left\{\bdPhi(\bdtheta) 
\bdSigma^{(l+1)} \bdPhi(\bdtheta)^H\right\},
\end{equation}
As seen from \eqref{Obj_L4}, a closed-form update equation for $\bdtheta$ can not be obtained, but we can 
use numerical methods, for instance, gradient descent (GD) 
to update $\bdtheta$ in the $l$-th iteration. However, GD employs backtracking line search \cite{NoceWrig06} 
to adaptively select the step-size which requires constant evaluation 
of the objective function in \eqref{Obj_L4}. 
Thus, to reduce the computational complexity we adopt the following single-step update for 
$\bdtheta$ with a constant step-size as suggested in \cite{off_grid1,nU_Burst_Dai_2019}, i.e., 
\begin{equation}\label{EM_GR}
\bdtheta^{(l+1)}=\bdtheta^{l}-\frac{r_\theta}{100}\text{sign}
\left\{\nabla_{\bdtheta^l}\bdL_2\left(\bdtheta^l\right)\right\},
\end{equation}
where $r_\theta$ is the grid interval, and $\text{sign\{.\}}$ represent the signum function which has negligible computational 
complexity. The step size $r_\theta/100$ divides the grid interval into $100$ equal parts, thus in the worst case the true values may be obtained in less than 
$100$ iterations. Further, this step size ensures that the final direction 
mismatch error is less than $1\%$ of $r_\theta$ which for sufficiently small $r_\theta$ is negligible to have significant 
impact on the channel estimation error.

The $m^{th}$ term of the gradient 
$\nabla_{\bdtheta}\bdL_2(\bdtheta)$ is given by
\begin{align}
&\left[\nabla_{\bdtheta^l}\bdL_2\left(\bdtheta^l\right)\right]_m=\frac{\partial}{\partial \theta^l_m}
\bdL_2\left(\bdtheta^l\right)
\nonumber\\
&=2\alpha^{(l+1)}_1\Re\{\dot{\bda}^H(\theta^l_m)\bdX^H\bdX{\bda}(\theta^l_m)\}
+2\Re\{\dot{\bda}^H(\theta^l_m)\bdX^H\bdalpha^{(l+1)}_2\},
\end{align}
in which, $\alpha^{(l+1)}_1=|\mu^{(l+1)}_m|^2+\Sigma^{(l+1)}_{m,m}$, 
$\bdalpha^{(l+1)}_2=\bdX\sum_{n\neq m}\Sigma^{(l+1)}_{n,m}\bda(\theta^l_n)-\bdy^{(l+1)}_{\backslash m}
\left(\mu^{(l+1)}_m\right)^*$, 
and $\bdy^{(l+1)}_{\backslash m}=\bdy-\bdX\sum_{n\neq m}(\mu^{(l+1)}_n\bda(\theta^l_n))$. 
The scalar $\Sigma^{(l+1)}_{n,m}=\left[\bdSigma^{(l+1)}\right]_{n,m}$ and 
the vector $\dot{\bda}(\theta^l_m)=\frac{\partial}{\partial \theta^l_m}\bda(\theta^l)$ is computed 
from \eqref{beam_steer} for $m=1,2,\ldots,M$. 

This completes all the sequential updates required to estimate $\bdxi$ in the $(l+1)$-st iteration. The parameters in $\bdxi$ are repeatedly updated in the EM iterations until convergence. The overall EM-EP algorithm is summarized in Algorithm \ref{algo2}. 
\begin{algorithm}\label{algo2}
  \footnotesize
\DontPrintSemicolon
\SetKwInput{KwPara}{Parameters}
\SetKwFor{ForEach}{for each}{}{end}
\SetKwRepeat{Repeat}{repeat}{until}
\SetKw{Break}{break}
\KwIn{$\bdy$}
\KwPara{$\bdxi^{(0)}$,$\bdtheta^{(0)}$, $\mu_{2,m}=0$,
$\Sigma_{2,m}=10^2$, $p_{2,m}=0$, 
$p^F_{3,m}=0$ for $m=2,\ldots,M$, $p^R_{3,m}=0$ for $m=1,2,\ldots,M-1$.}
\tcc{EM-EP run}
\ForEach{$l = \{0,1,2,\ldots,n_{EM}-1\}$}
{
\begin{enumerate}
\item Given $\bdxi^{l}$ and $\bdtheta^{l}$ run the EP algorithm 
described in\\ Algorithm \ref{algo1} to generate $\bdmu^{(l+1)}$, $\bdSigma^{(l+1)}$, and $\bdp^{(l+1)}$.
\item {Check for convergence:}\\
\If{$\frac{||\bdmu^{(l+1)}-\bdmu^{l}||}{||\bdmu^{l}||}<\epsilon_{EM}$}{\quad break;}
\item Use $\bdmu^{(l+1)}$, 
$\bdSigma^{(l+1)}$, and $\bdp^{(l+1)}$ to update $\tau_{10}$, $\tau_{01}$, \\ $\bdgamma$, 
$\eta$, and $\bdtheta$ using \eqref{updated_p01}, \eqref{updated_p10}, 
\eqref{updated_gamma_m}, \eqref{updated_eta}, and \eqref{EM_GR}, 
respectively.
\end{enumerate}
}
\KwOut{$\hat{\bdh}=\bdA(\bdtheta^{(l+1)})\bdmu^{(l+1)}$}
\caption{Overall EM-EP Algorithm}
\end{algorithm}
\begin{figure*}[ht]
    \begin{minipage}{0.50\textwidth}
        \centering
				\includegraphics[width=1.0\textwidth,height=0.6\textwidth]{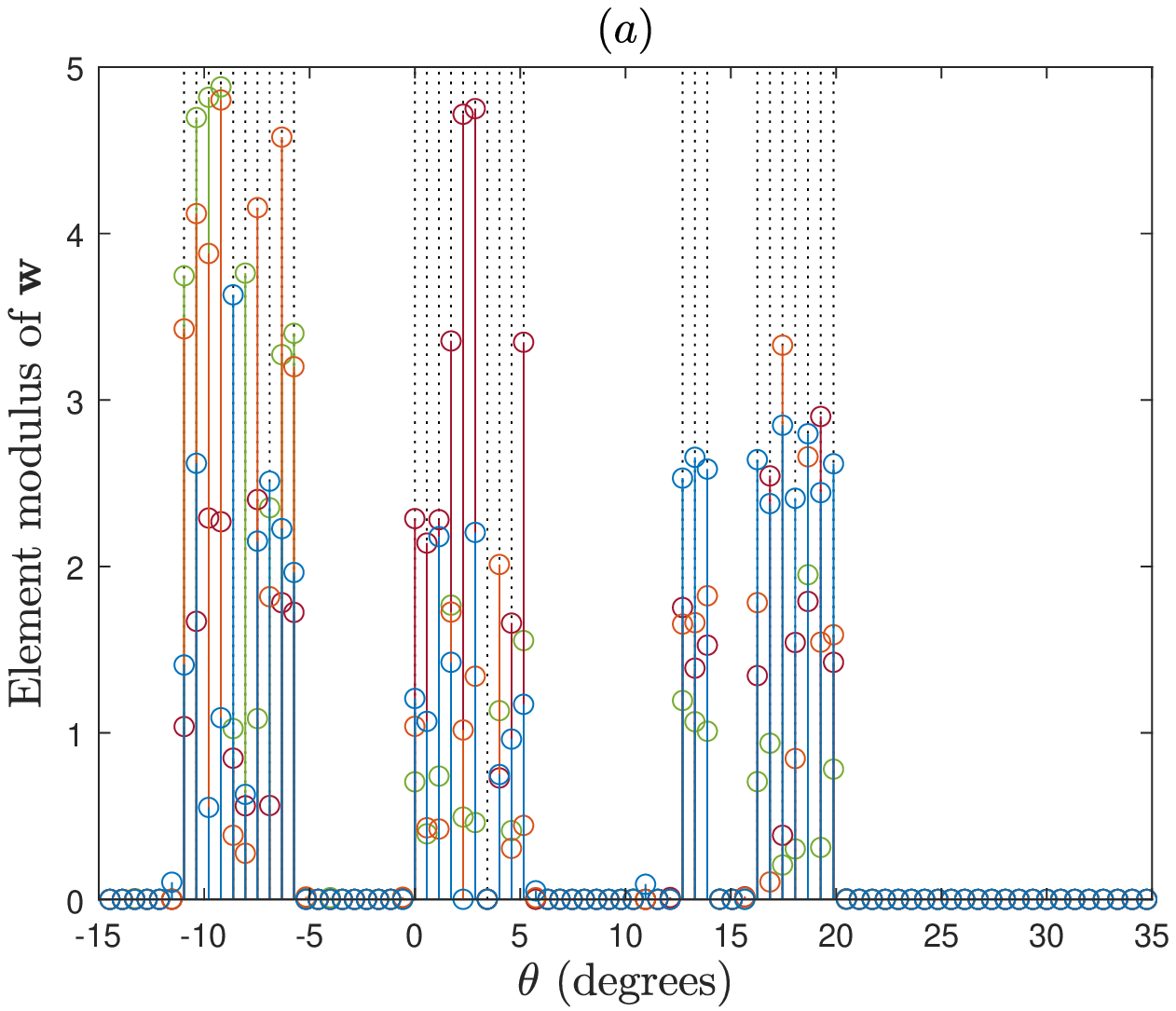}
    \end{minipage}
    \begin{minipage}{0.50\textwidth}
        \centering
\includegraphics[width=1.0\textwidth,height=0.6\textwidth]{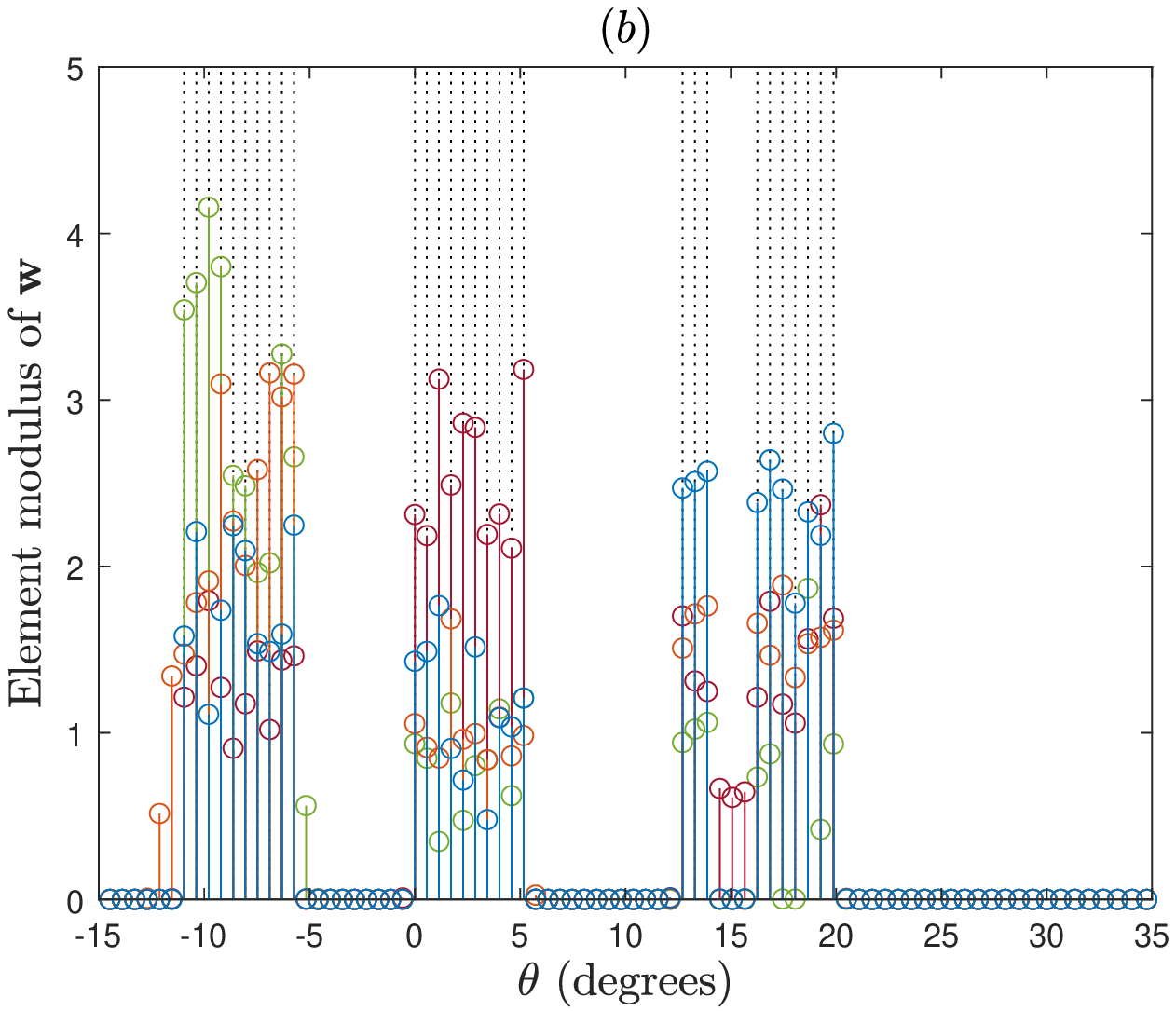}
		\end{minipage}
		\\
    \begin{minipage}{0.50\textwidth}
        \centering
\includegraphics[width=1.0\textwidth,height=0.6\textwidth]{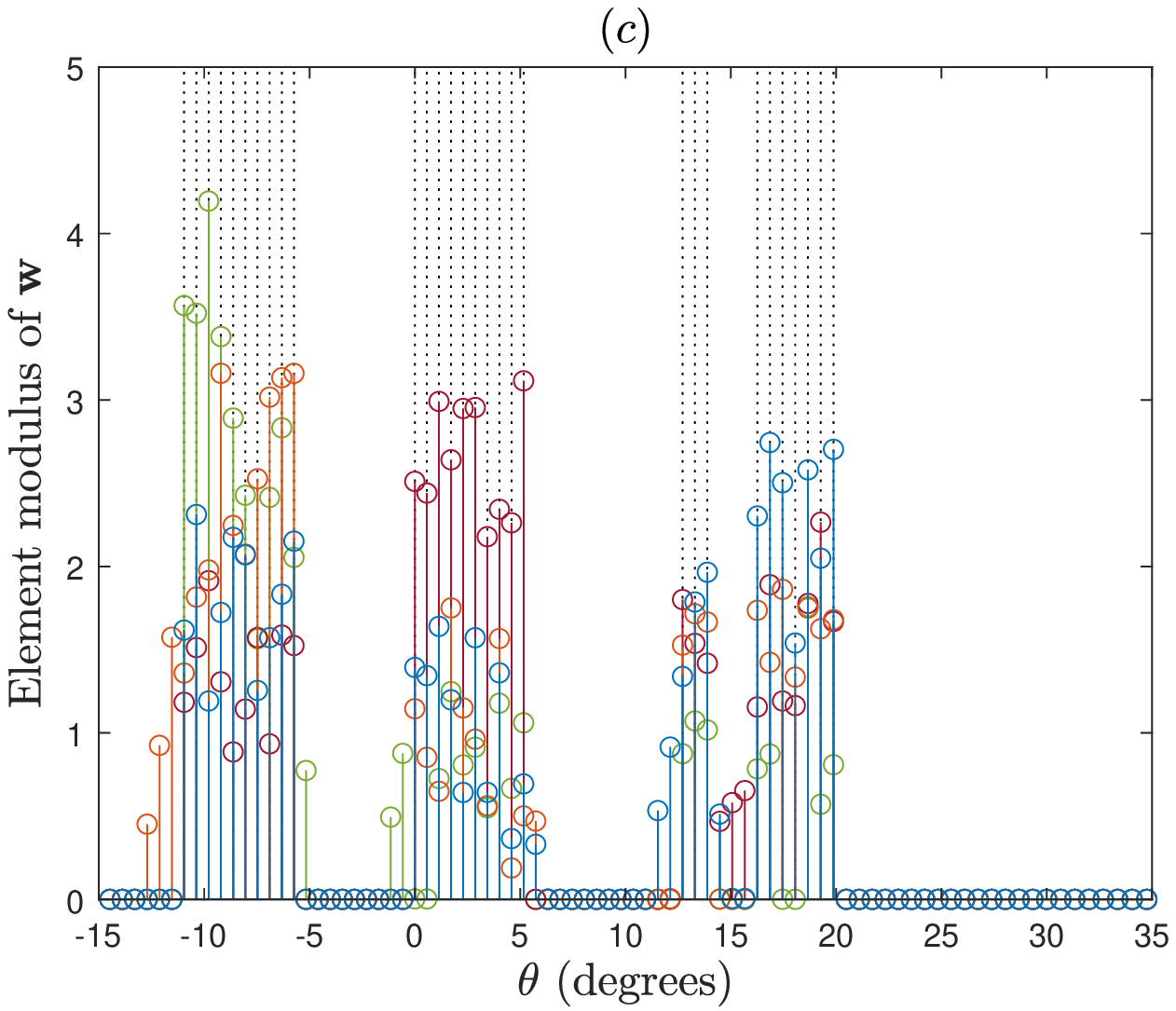}
		\end{minipage}
    \begin{minipage}{0.50\textwidth}
        \centering
     	\includegraphics[width=1.0\textwidth,height=0.6\textwidth]{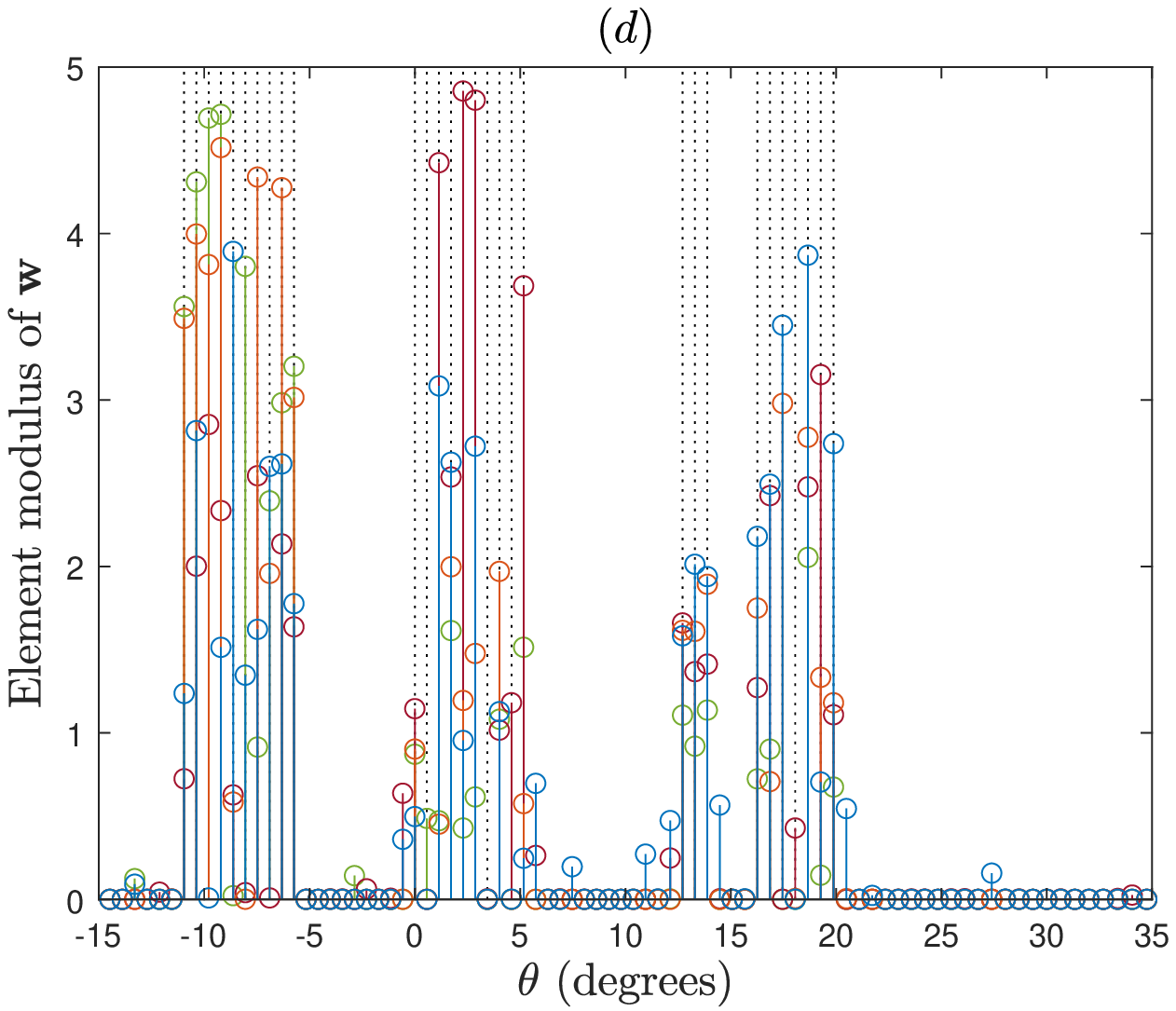}
    \end{minipage}
		\\
    \begin{minipage}{0.50\textwidth}
        \centering
\includegraphics[width=1.0\textwidth,height=0.6\textwidth]{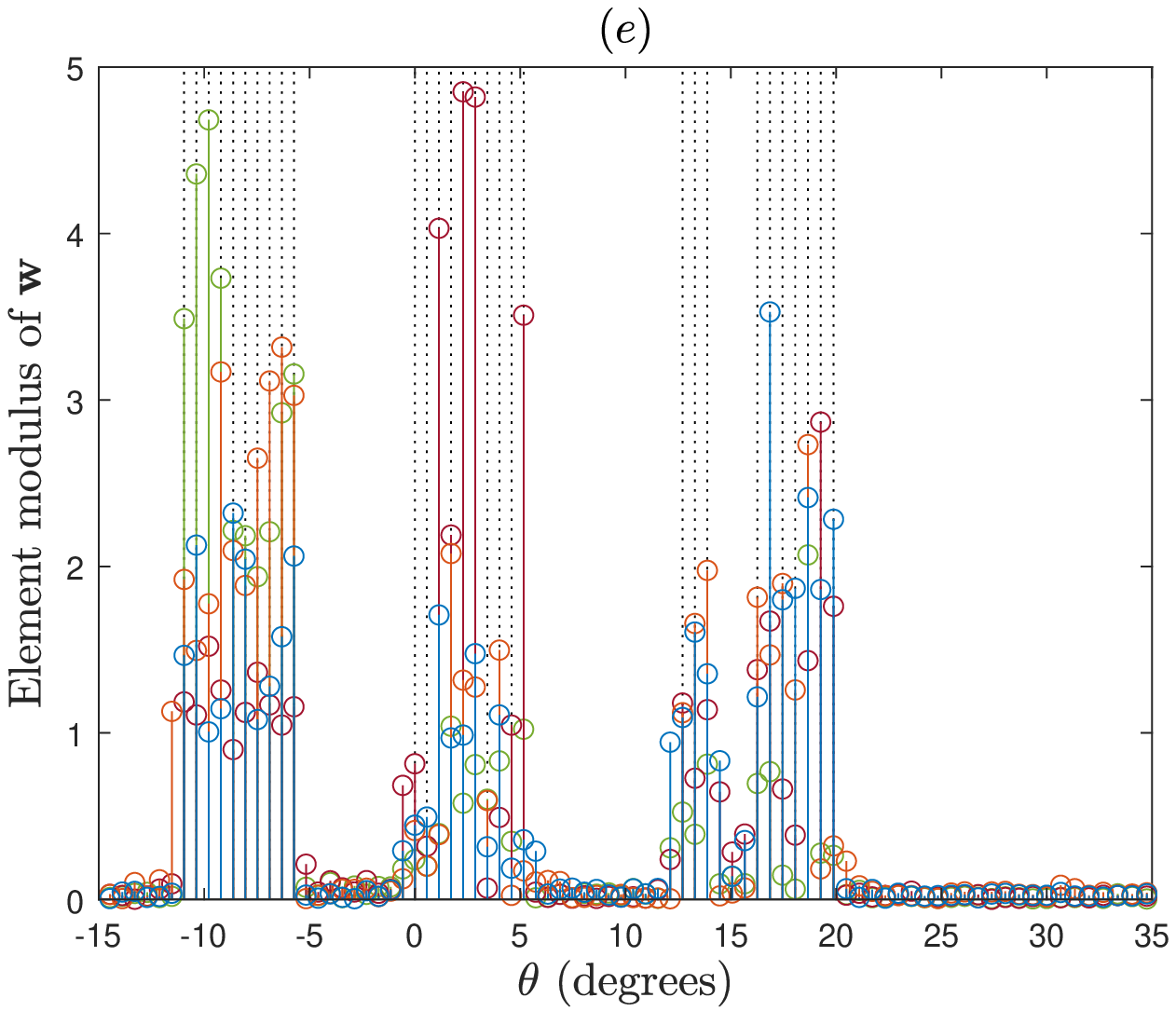}
		\end{minipage}
    \begin{minipage}{0.50\textwidth}
        \centering
\includegraphics[width=1.0\textwidth,height=0.6\textwidth]{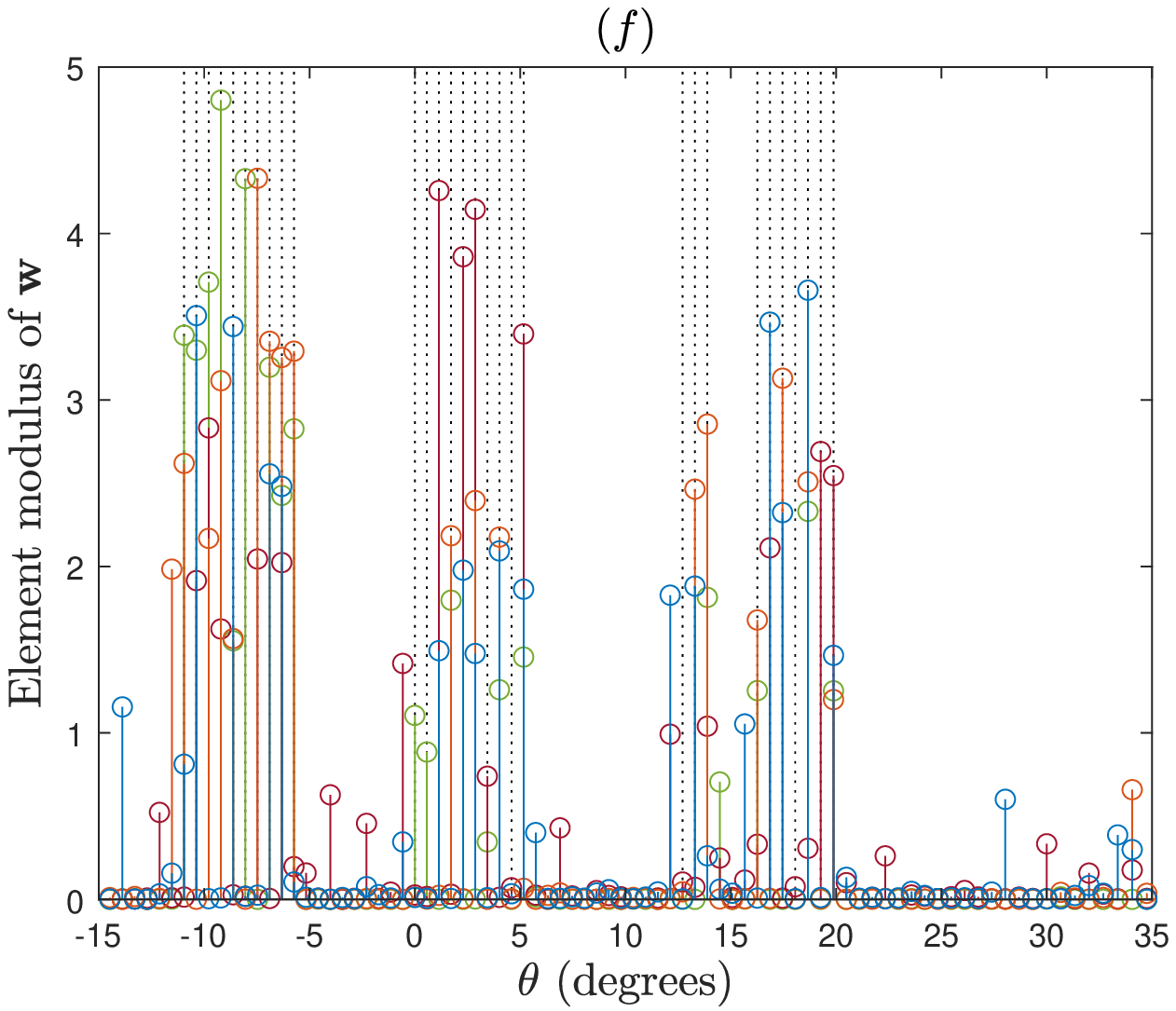}
		\end{minipage}

		\caption{Magnitude of the elements in $\bdw$ for four independent trials with 
		$G=128$, $M=200$, $N=48$, $L_s=3$, $L_p=10$, and $SNR=10$ dB, and for $(a)$ EM-EP, $(b)$ SuRe-CSBL, 
		$(c)$ S-TCS, $(d)$ EM-EP-B, $(e)$ PC-VB, $(f)$ EM-BG-GAMP. The dotted lines indicate locations of the true AoDs.}
		\label{fig:sparsity_recovery}
\end{figure*}

\subsection{Computational Complexity of EM-EP algorithm}

The computational complexity of the proposed EP algorithm per iteration is dominated 
by \eqref{Eq_bdV} and \eqref{Eq_bdm} which can be solved in $O(NM^2)$ 
computations. This complexity is the same as that of the EP algorithm 
proposed in \cite{Hernandez_2015}. For the 
EM part of the algorithm, the dominant terms include the 
update of $\eta$ by \eqref{updated_eta} which takes $O(NM^2)$ computations, 
and the update of $\bdtheta$ by \eqref{EM_GR} which takes $O(GNM)$ computations.
Since $M$ is usually greater than $G$, the complexity of the proposed EM-EP 
algorithm is $O(NM^2)$ per iteration which is the same as that of the off-grid SBL 
algorithm proposed in \cite{off_grid1}.

\section{Simulation Results}\label{sim_results}
In this section, we investigate the performance of the proposed EM-EP algorithm 
for massive MIMO channel estimation.
We consider a single-cell where 
a BS equipped with a ULA has $G$ antennas and transmits $N$ pilot symbols 
to a reference user. The elements in the pilot matrix $\bdX$ are 
selected from a circularly symmetric complex Gaussian distribution with unit variance, 
and the DL channel $\bdh$ between the BS and the user is generated using the 
$3$GPP spatial channel model \cite{SCM_2012} with urban-micro cell environment. 
We assume that each channel realization is composed of $L_s$ scatterers 
with AoDs 
randomly located in the interval $[-90^{\text{o}},90^{\text{o}}]$, and each scatterer 
has $L_p$ paths with the AoDs randomly generated and concentrated in an 
angular spread denoted by $A$. 
Unless stated otherwise, the AoDs 
of all the paths in a channel realization are continuous-valued variables 
and thus may not lie on the assumed angular grid.
The DL channel 
frequency is selected as $2.17$ GHz and the spacing between adjacent antennas in the ULA is 
set as $d=\frac{c}{2f_0}$ where $c$ is the speed of light and $f_0=2$ GHz. 
\begin{figure*}[ht]
    \begin{minipage}{0.48\textwidth}
     	\includegraphics[width=0.99\textwidth]{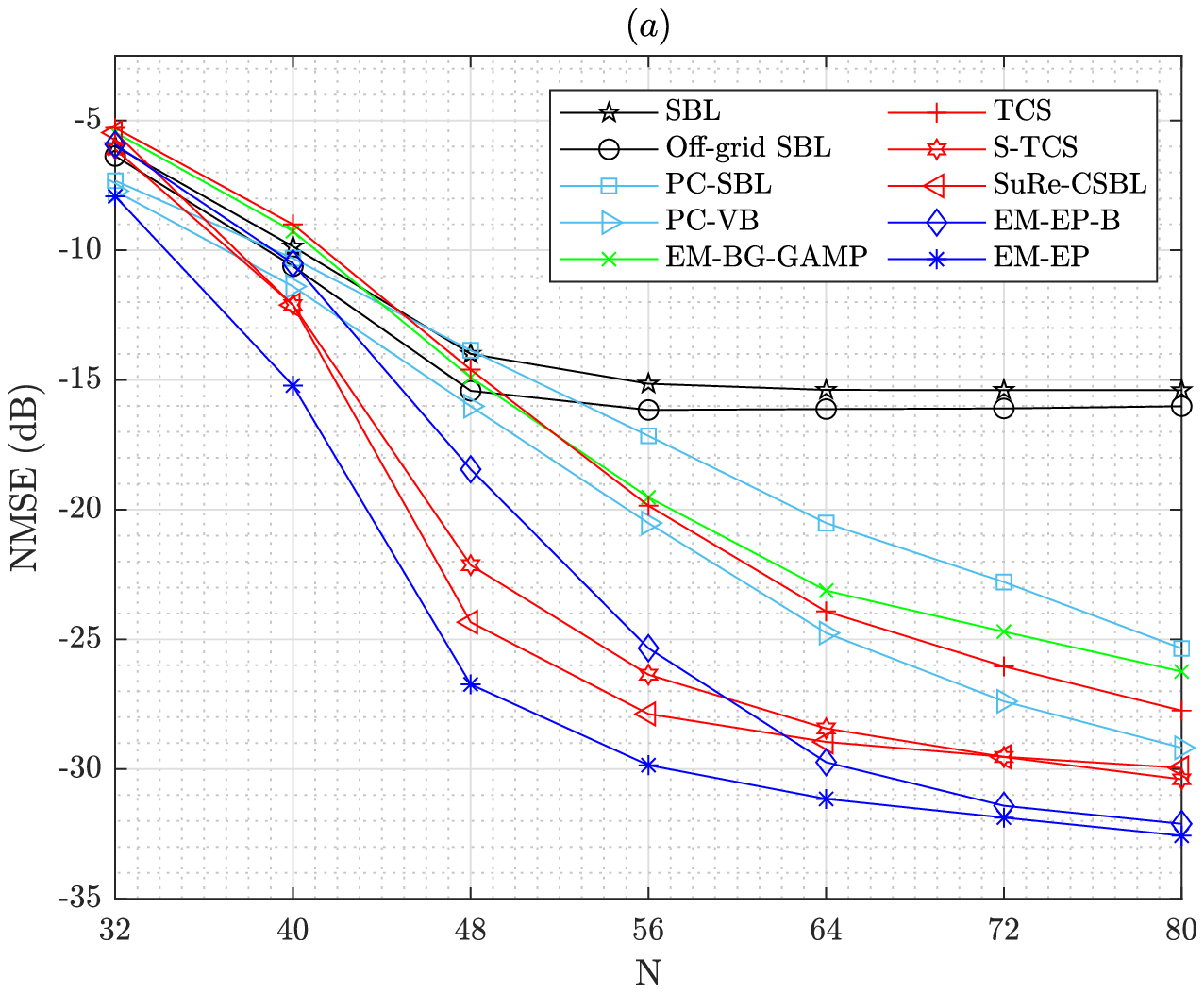}
    \end{minipage}\hspace{.01\linewidth}
    \begin{minipage}{0.48\textwidth}
\includegraphics[width=0.99\textwidth]{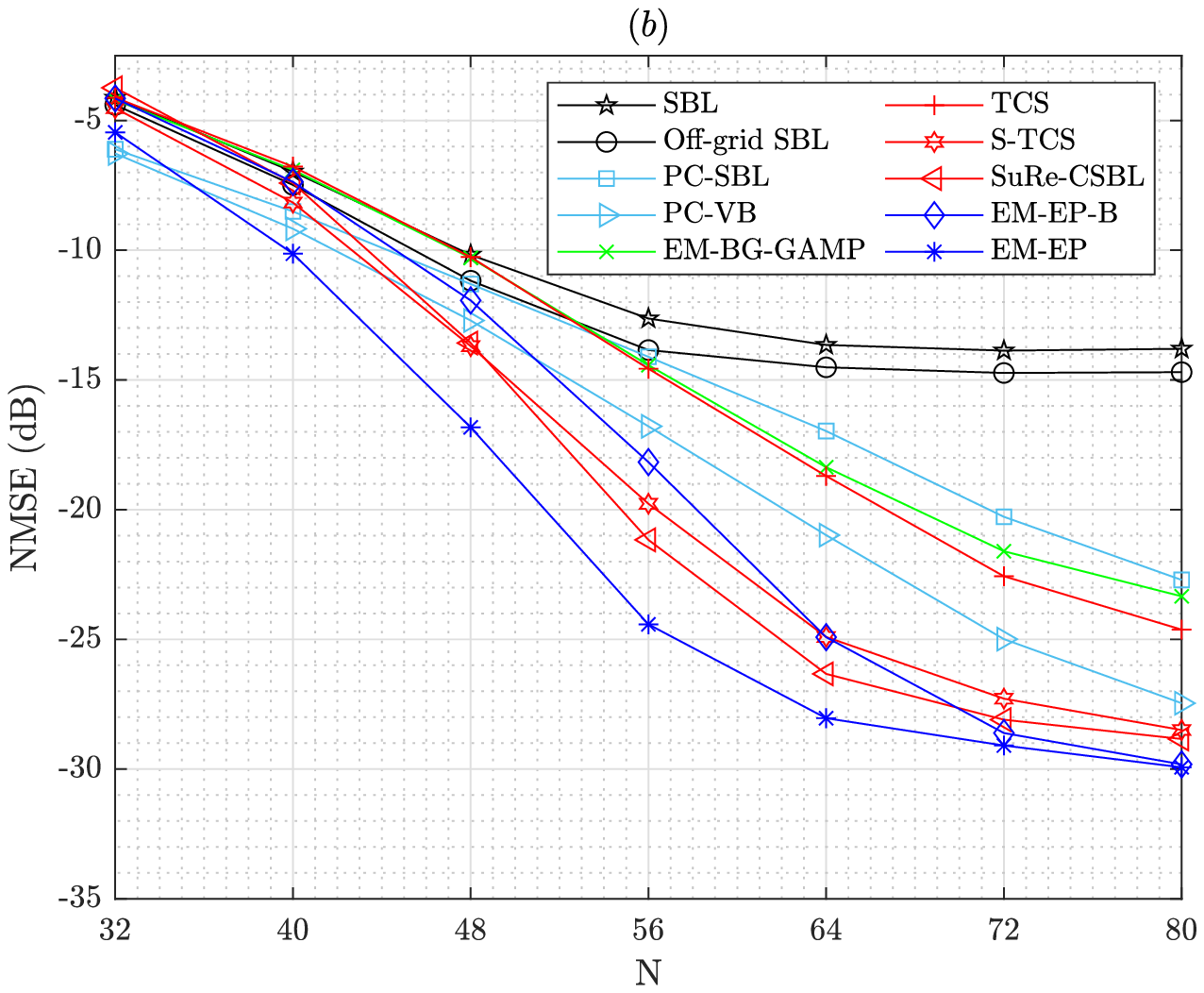}
		\end{minipage}
		\caption{Channel estimation error vs. number of pilot symbols $N$ with parameters 
		$G=128$, $M=200$, SNR=$10$ dB, and for $(a)$ $L_s=3$, $L_p=10$, and $A=10^\text{o}$,
		$(b)$ $L_s=4$, $L_p=10$, and $A=10^\text{o}$.}
		\label{fig:Error_vs_N}
\end{figure*}
\begin{figure*}[ht]
    \begin{minipage}{0.48\textwidth}
     	\includegraphics[width=0.99\textwidth]{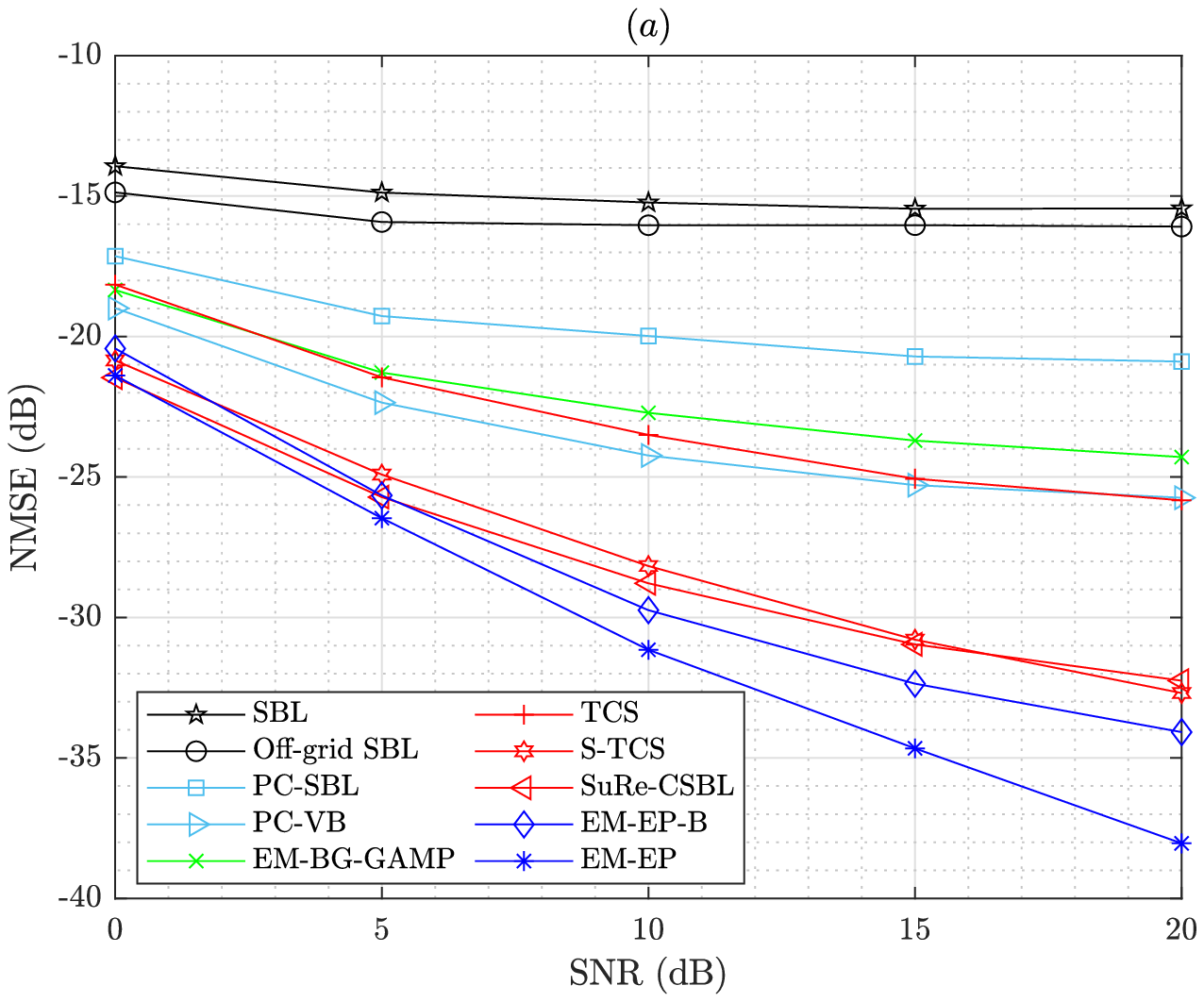}
    \end{minipage}\hspace{.01\linewidth}
    \begin{minipage}{0.48\textwidth}
\includegraphics[width=0.99\textwidth]{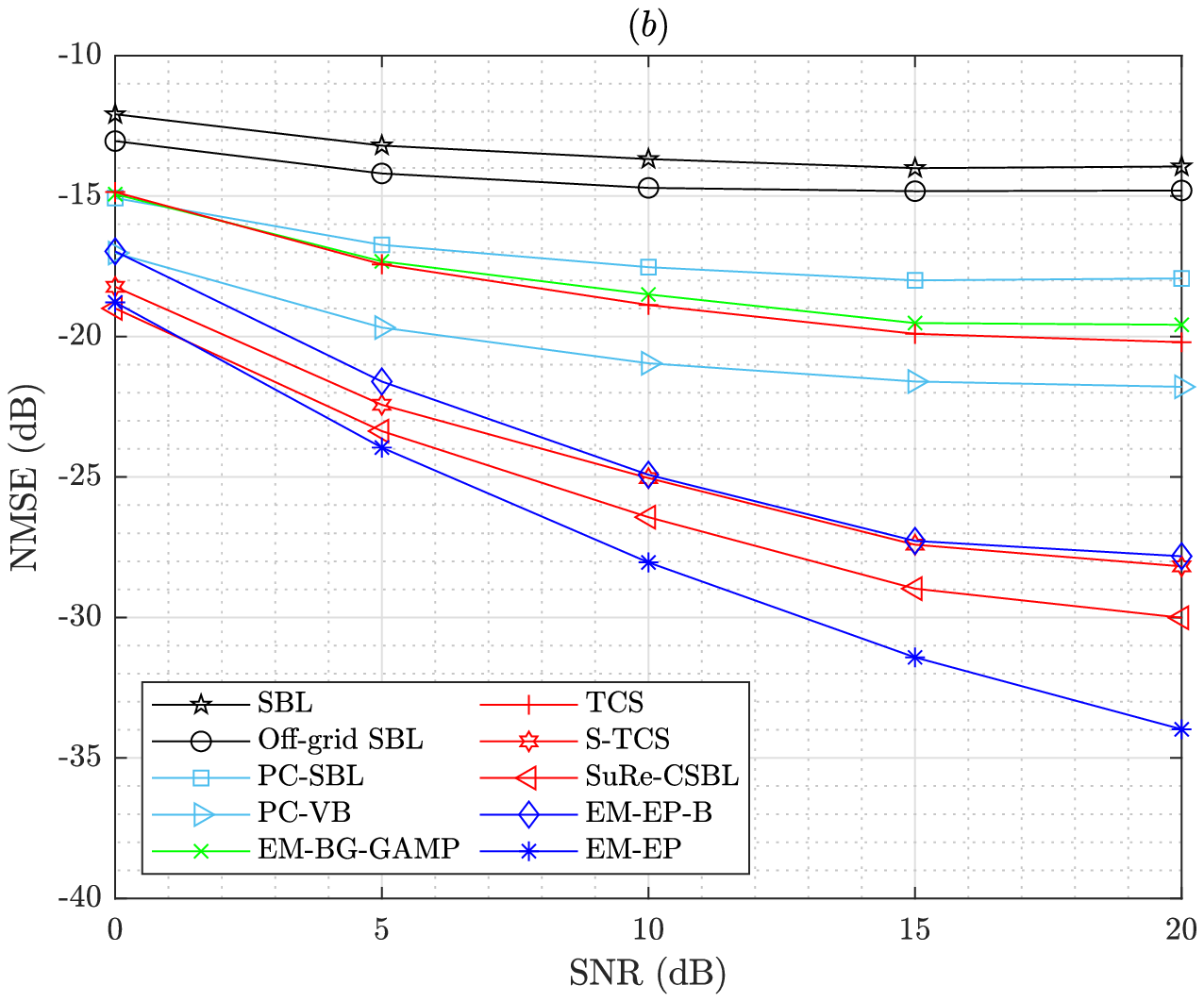}
		\end{minipage}
		\caption{Channel estimation error vs. SNR (dB) with parameters 
		$G=128$, $M=200$, $N=64$, and for $(a)$ $L_s=3$, $L_p=10$, and $A=10^\text{o}$,
		$(b)$ $L_s=4$, $L_p=10$, and $A=10^\text{o}$.}
		\label{fig:Error_vs_SNR}
\end{figure*}


In order to compare our algorithm with 
the EP algorithm proposed in \cite{Hernandez_2015}, 
we need to extend this algorithm.
In \cite{Hernandez_2015} 
the authors modeled the elements of the support (latent) vector $\bdz$ with an iid Bernoulli prior distribution having a 
parameter $p_0$ which, along with the other model parameters, is assumed to be known. 
To apply their approach to the problem under consideration here, 
we need to estimate these parameters. Therefore we extend the method in 
\cite{Hernandez_2015}
with the EM algorithm as discussed in section \ref{sys-model}
and refer to the resulting algorithm as EM-EP-B.
More specifically, in the {(l+1)-st} iteration of 
EM-EP-B algorithm, $p_0$ is updated according to 
$p^{(l+1)}_0=\frac{1}{M}\sum^M_{m=1}\sigma(p^{(l+1)}_m)$.
Moreover, the other model parameters, i.e., $\eta$, $\gamma_m$, 
and (to integrate grid refining) $\bdtheta$ are updated 
using our results in \eqref{updated_gamma_m}, \eqref{updated_eta}, and \eqref{EM_GR} 
from section \ref{EM_part}. 
 
We also show the performances of 
SBL \cite{SBL_3}, Off-grid SBL \cite{off_grid1}, 
PC-SBL \cite{PC_SBL1}, PC-VB \cite{nU_Burst_Dai_2019}, 
EM-BG-GAMP \cite{Vila_EM-BG-GAMP}, TCS \cite{TCS_paper}, S-TCS \cite{S_TCS_paper}, 
and SuRe-CSBL \cite{Super_Markov_1}.
For Off-grid SBL, PC-SBL, PC-VB, SuRe-CSBL, EM-EP-B,
and EM-EP algorithms, the dictionary $\bdA(\bdtheta)$ is initialized 
to be a (partial) DFT matrix.
For the other algorithms, however, $\bdA(\bdtheta)$ is the fixed 
DFT matrix as required for the derivation of
the algorithms and state evolution analysis\footnote{For consistency, to initialize
EM-EP-B, we set $p^{(0)}_0=\lambda^{(0)}$ whereas the other hyperparameters and the termination condition were set the same as those 
for EM-EP.
To compare our results with TCS and S-TCS, all the hyperparameters were updated 
using the EM update equations from \cite{Vila_EM-BG-GAMP} except for the transition probabilities for S-TCS which were updated 
using the posterior means in \eqref{updated_p01} and \eqref{updated_p10}.}. 
In all the experiments, we initialized the EM-EP algorithm with 
$\lambda^{(0)}=0.3$, $\tau_{01}^{(0)}=0.1$, 
$\tau_{10}^{(0)}=\frac{\lambda^{(0)}}{1-\lambda^{(0)}}\tau^{(0)}_{01}$, 
$\eta^{(0)}=\gamma^{(0)}_m=\left(\frac{||\bdy||^2}{(SNR^{(0)}+1)N}\right)^{-1}$ 
{with $SNR^{(0)}=100$}, and 
$\theta^{(0)}_m=\text{sin}^{-1}\left(-1+\frac{2m}{M}\right)$ 
for $m=1,2,\ldots, M$ 
{as in \cite{Vila_EM-BG-GAMP, Super_Markov_1}.} 
The maximum iterations of EM and EP algorithms are set as $n_{EP}=n_{EM}=100$ 
and the tolerance coefficients are selected to be $\epsilon_{EP}=\epsilon_{EM}=10^{-4}$. 
The channel estimation error is computed by using the following normalized 
mean-squared-error (NMSE), 
\begin{equation}
\text{NMSE (dB)}=10\log_{10} \frac{\Exp[||\hat{\bdh}-\bdh||^2]}{\Exp[||\bdh||^2]},
\end{equation}  
in which $\hat{\bdh}$ is the channel estimate.

In Fig. \ref{fig:sparsity_recovery} we investigate the performance of the selected channel estimation algorithms for 
recovering the sparse vector $\bdw$ with non-uniform burst sparsity\footref{foot_1}. 
We consider a BS with $G=128$ antennas 
transmitting $N=48$ pilot symbols to the user with $SNR=10$ dB. 
The physical channel between the BS and the user has $L_s=3$ 
scatterers with $L_p=10$ paths per scatterer. The channel estimators assume a fixed uniformly-spaced angular grid with 
$\theta_m=\text{sin}^{-1}\left(-1+\frac{2m}{M}\right)$ for $m=1,2,\ldots, M$ with $M=200$, and the physical AoDs 
corresponding to the three non-zeros clusters 
are assumed to be located on the grid points at 
$m=81,82,\ldots,90$,$100,101,\ldots,109$, $122,123,124,128,129,\ldots,134$. We get the following observations from Fig. 
\ref{fig:sparsity_recovery}. 
Firstly, when the non-zero clusters are closely located as shown by the 
dotted lines in Fig. \ref{fig:sparsity_recovery}, the algorithms such as PC-VB 
which tune each coefficient based on the nearest neighbor,
exhibit a performance loss due to the leakage of energy into the bins 
between the adjacent clusters.
For instance, observe the energy leakage around $-3^\deg$, $8^\deg$, and $15^\deg$ in 
Fig. \ref{fig:sparsity_recovery} (e). Secondly, the algorithms such as EM-EP-B and EM-BG-GAMP which aim to recover the 
coefficients individually result in outliers at random positions far away 
from the true AoDs. This effect, when pronounced as in the case of EM-BG-GAMP,
causes significant performance loss. 
Thirdly, SuRe-CSBL and S-TCS which employ a Markov prior on the support vector $\bdz$ eliminate the outliers, but suffer from significant leakage of energy into 
the bins near the clusters true AoDs. Finally, 
our proposed EM-EP algorithm eliminates the 
leakage of energy as well as the occurrence of outliers, and much more
accurately represents the channel.

\begin{figure*}[ht]
    \begin{minipage}{0.48\textwidth}
     	\includegraphics[width=0.99\textwidth]{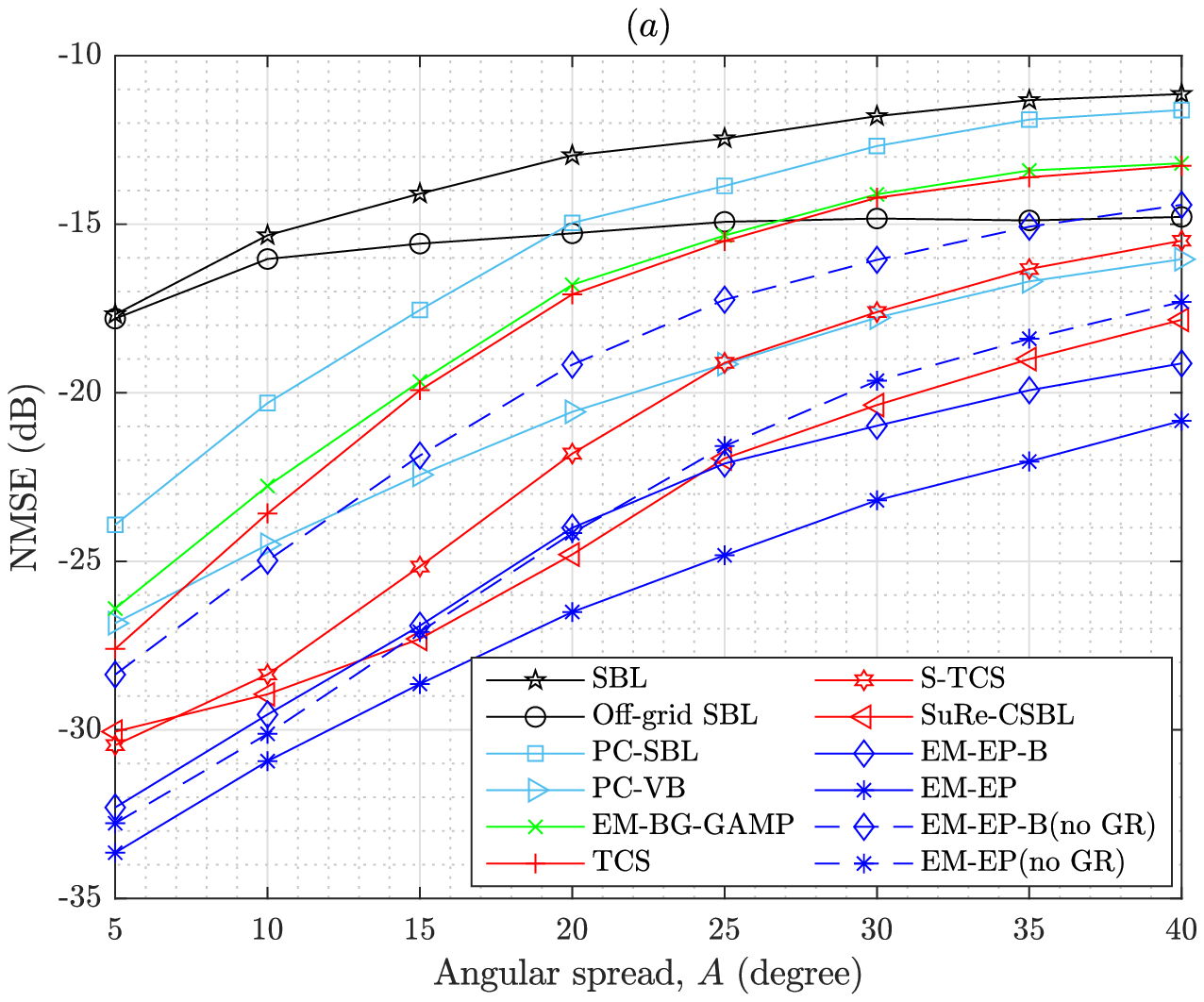}
    \end{minipage}
    \begin{minipage}{0.48\textwidth}
\includegraphics[width=0.99\textwidth]{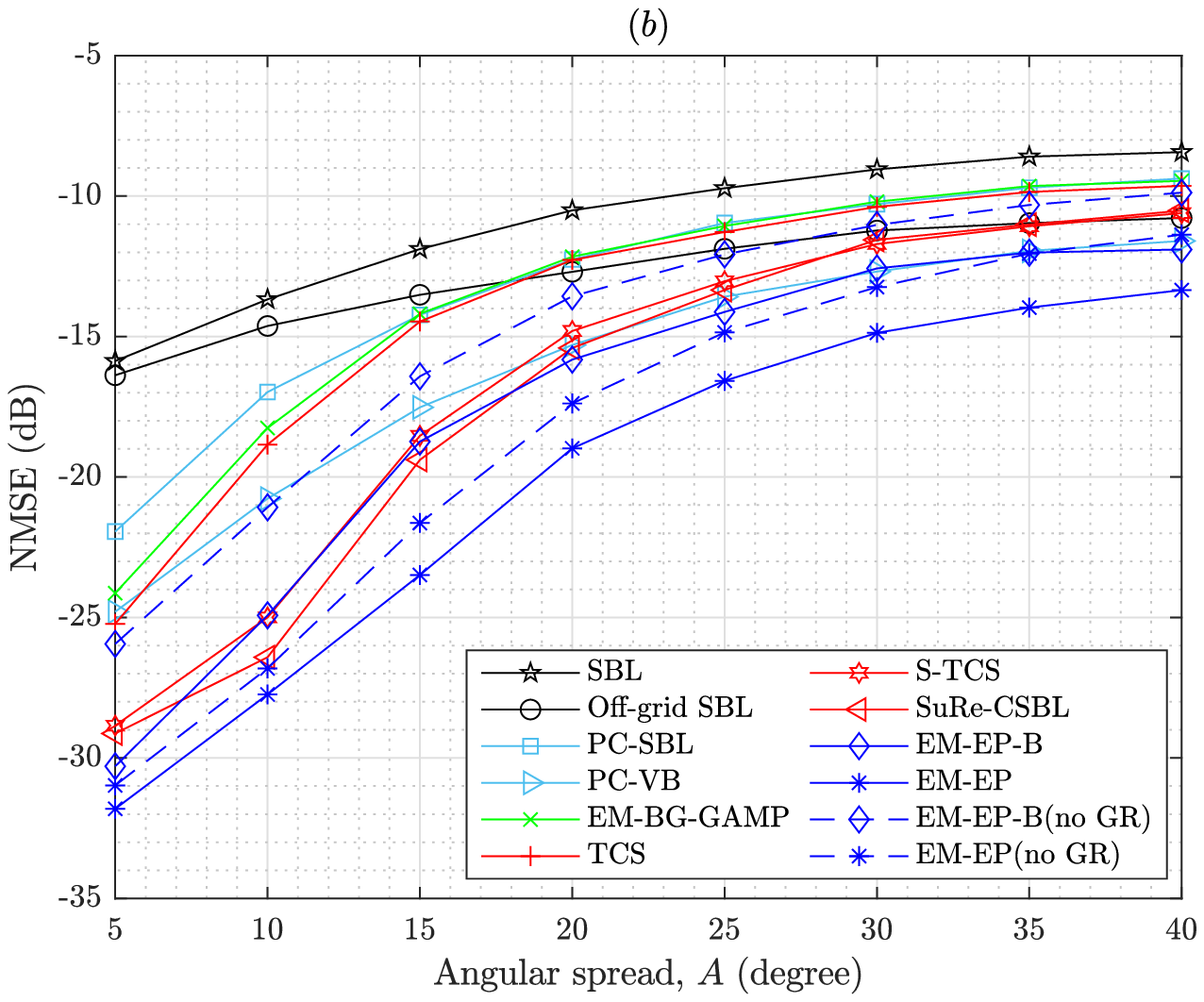}
		\end{minipage}
		\caption{Channel estimation error vs. Angular spread $A$ with parameters 
		$G=128$, $M=200$, $N=64$, and for $(a)$ $L_s=3$, $L_p=10$, and $SNR=10$ dB,
		$(b)$ $L_s=4$, $L_p=10$, and $SNR=10$ dB.}
		\label{fig:Error_vs_A}
\end{figure*}
\begin{figure*}[ht]
    \begin{minipage}{0.48\textwidth}
     	\includegraphics[width=0.99\textwidth]{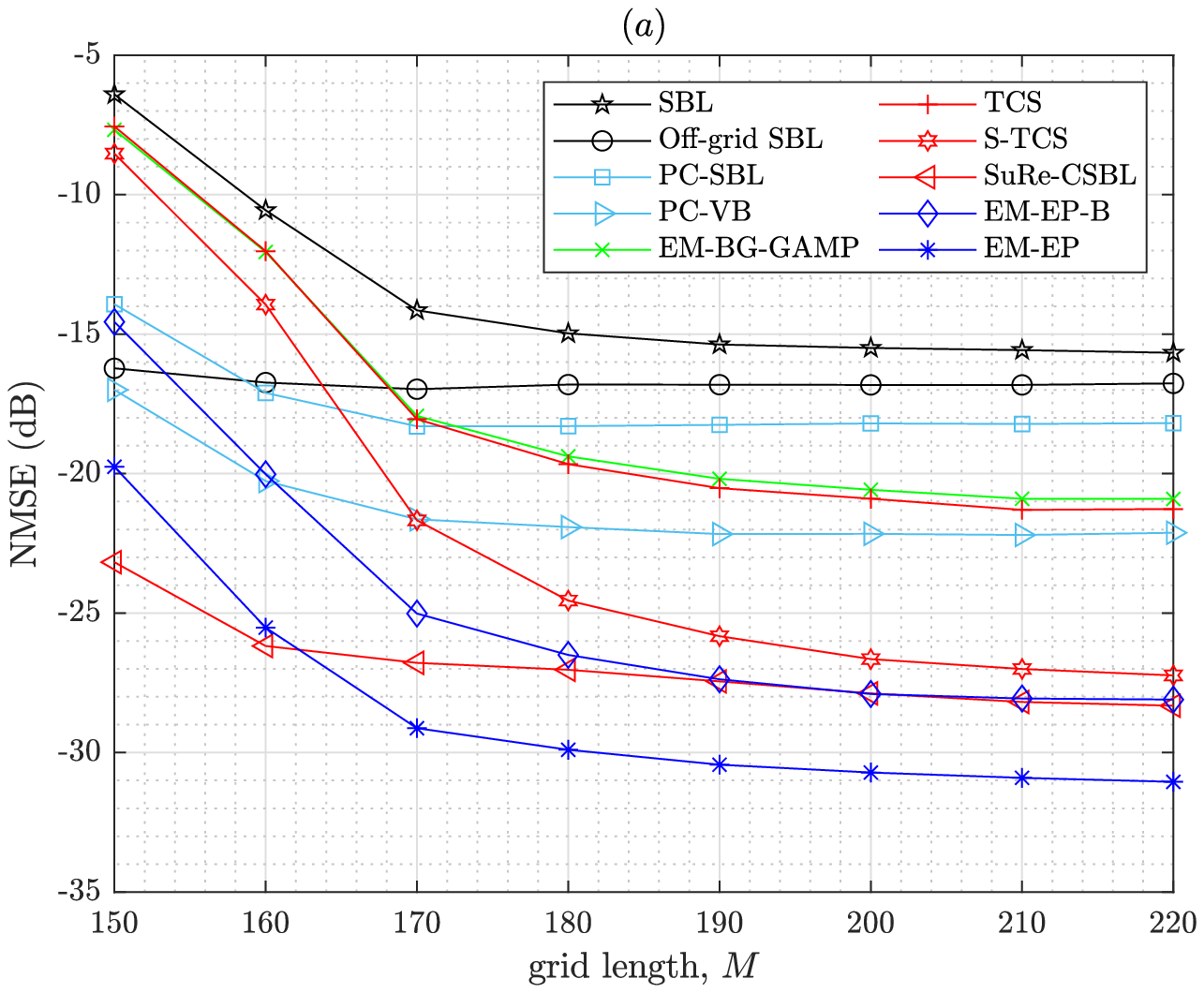}
    \end{minipage}
    \begin{minipage}{0.48\textwidth}
\includegraphics[width=0.99\textwidth]{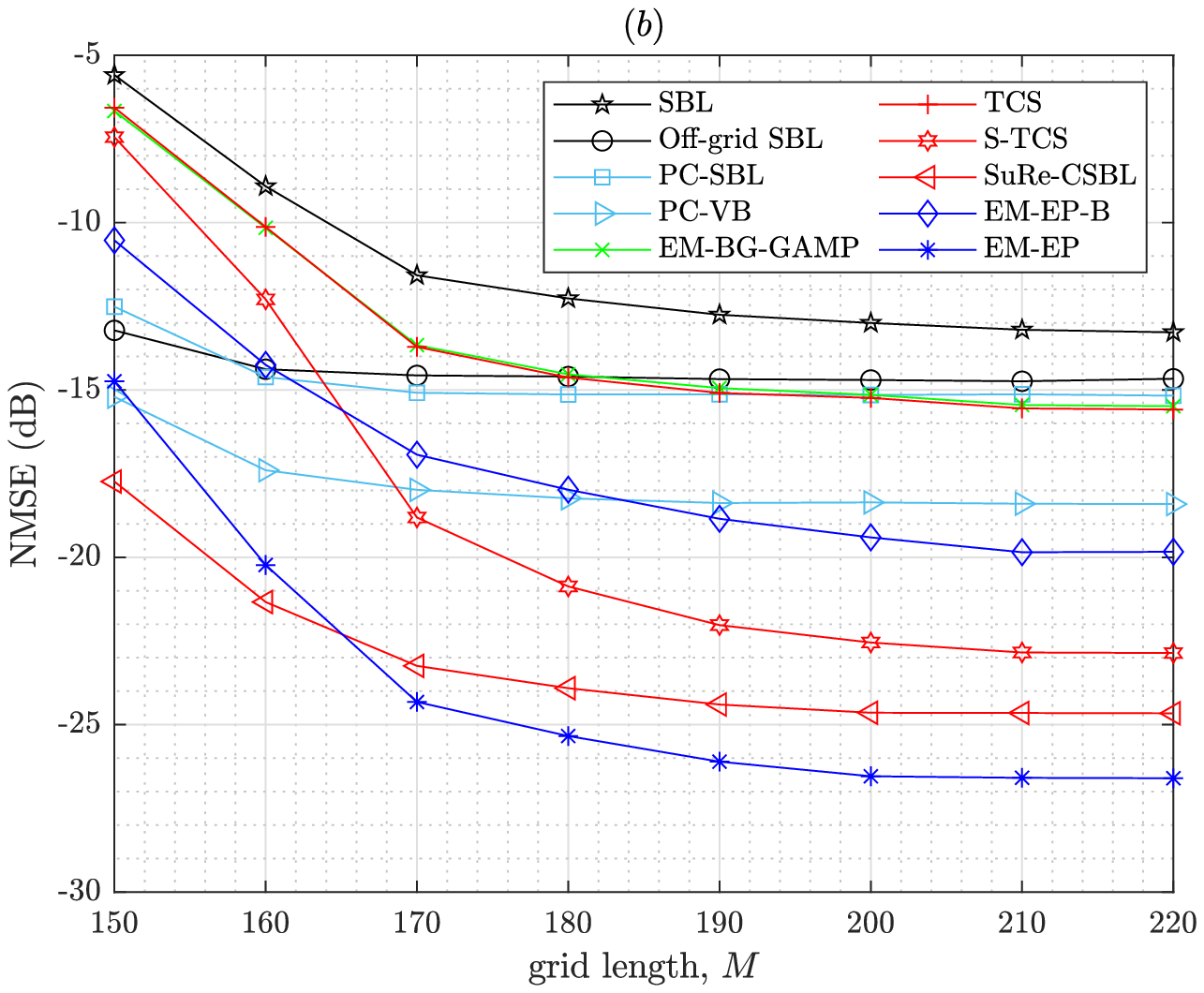}
		\end{minipage}
		\caption{Channel estimation error vs. grid length $M$ with parameters 
		$G=150$, $N=64$, $SNR=10$ dB, and for $(a)$ $L_s=3$, $L_p=10$, and $A=10^\deg$,
		$(b)$ $L_s=4$, $L_p=10$, and $A=10^\deg$.}
		\label{fig:Error_vs_M}
\end{figure*}

Fig. \ref{fig:Error_vs_N} shows the channel estimation error versus the number of pilot symbols $N$ 
for the selected channel estimation schemes. 
We consider the massive MIMO channel with 
$L_s=3$ or $4$ scatterers and $L_p=10$ paths per scatterer. The AoDs for all the paths 
are randomly generated continuous-valued parameters with no on-grid assumption as before, and all the paths 
per scatterer are concentrated in an angular spread $A=10^\deg$. 
We observe that in both cases shown in Figs. \ref{fig:Error_vs_N}(a) and 
\ref{fig:Error_vs_N}(b),
the performance of the algorithms improve with $N$ and EM-EP significantly
outperforms all the algorithms. The channel has more paths in case of $L_s=4$ in 
Fig. \ref{fig:Error_vs_N}(b)
and thus larger values of $N$ are required to reach the same level of performance. 
SBL, Off-grid SBL, EM-BG-GAMP, and EM-EP-B 
aim to recover the coefficients individually and hence their 
performance is degraded due to the occurrence of 
outliers in the angular domain. Compared to SBL and Off-grid SBL which assume an iid complex Gaussian prior on $\bdw$, 
EM-BG-GAMP, TCS, and EM-EP-B assume 
an iid Bernoulli-Gaussian (BG) prior where the level of sparsity in $\bdw$ 
is directly adjusted by the weight of the Bernoulli component. 
This weight determines the 
fraction of coefficients that are a priori set to zero. 
Thus, EM-BG-GAMP and TCS perform better than the 
SBL-based algorithms. On the other hand,
EM-EP-B includes the correlations in $\bdw$ by using $\bdSigma$ 
in its estimation of the posterior distribution and also performs 
grid refining to learn the dictionary. Therefore EM-EP-B outperforms 
both EM-BG-GAMP and TCS. 
PC-SBL and PC-VB aim to recover each coefficient in $\bdw$ according 
to its nearest neighbor. PC-SBL uses an SBL-based algorithm and tunes the precision of each coefficient 
according to the precisions of its immediate neighbors 
but using a sub-optimal solution. 
PC-VB avoids this sub-optimality 
by linking a support vector with a multinoulli prior to every coefficient and 
using a variational Bayes (VB) \cite{VB_algo} based algorithm.
Hence, PC-VB performs better than PC-SBL, but its performance suffers due to 
the leakage of energy when multiple non-zeros 
clusters are closely located.
Performance of PC-VB is inferior to that of EM-EP-B.
Due to its dependence on the VB method, 
PC-VB may approximate the true distribution locally around one of its several 
sub-optimal modes, 
whereas EM-EP-B employs the EP method which approximates the true distribution globally over a wider support and thus results in a better performance \cite{Bishop_2006}. 
Finally in contrast to S-TCS and SuRe-CSBL, EM-EP takes into account the
correlation in $\bdw$ thereby outperforming the former two algorithms. 


Fig. \ref{fig:Error_vs_SNR} shows the channel estimation error versus SNR for the selected 
algorithms. We consider the same scenario as in Fig. \ref{fig:Error_vs_N} except that the number of pilot 
symbols is now fixed to $N=64$. We observe that 
the performance of the algorithms improves with SNR and the proposed EM-EP algorithm has the best performance of all the schemes.
In case of $L_s=4$ scatterers the channel has more paths 
and therefore has more chances of having {non-equal}
size clusters.
Therefore in this case the performance of EM-EP-B which aims to recover 
{the coefficients individually}
deteriorates and is worse than that of SuRe-CSBL.
Fig. \ref{fig:Error_vs_SNR} also shows that while
the performance of all the methods reaches a floor at some value of SNR 
(This is more evident in Fig. \ref{fig:Error_vs_SNR}(b).),
the proposed EM-EP continues to improve with SNR.


Fig. \ref{fig:Error_vs_A} shows the channel estimation error for different values of
the angular spread $A$. We consider two cases of 
$L_s=3$ and $L_s=4$ scatterers as before and with $G=128$, $M=200$, and $N=64$. 
The SNR value is fixed to $10$ dB. 
As $A$ increases severe {non-equal size burst sparsity}
may exist with 
isolated paths, and thus as observed from Fig. \ref{fig:Error_vs_A} 
the performance of the 
algorithms degrades accordingly. 
For a fixed $A$, such {non-equal size burst sparsity becomes more intense}
when the channel has more 
paths as in case (b), and hence the channel estimation errors are relatively higher. 
However, in both cases the EP-based algorithms show significant gains in performance, 
and the proposed EM-EP algorithm outperforms all the algorithms. 
In Fig. \ref{fig:Error_vs_A}  we also show the performance of EM-EP when no 
grid refining is performed, i.e., {no optimization over AoDs $\bdtheta$,} denoted in
Fig. \ref{fig:Error_vs_A} as EM-EP(no-GR). 
It can be seen that EM-EP(no-GR) performs better than most of the other algorithms
due to the use of the EP method and taking into account the correlation in $\bdw$.

In Fig. \ref{fig:Error_vs_M} we examine the effect of varying the grid length $M$ on the channel estimation performance of the algorithms. Consider the channel with $L_s=3$ or $4$ scatterer where the BS has 
$G=150$ antennas, the number of pilot symbols are fixed to $N=64$, 
angular spread is selected to be $A=10^\deg$, and the SNR is set to $10$ dB. 
It is observed that in both cases shown in Fig. \ref{fig:Error_vs_M} the performance of the algorithms improve with $M$, and 
our proposed EM-EP algorithm outperforms all the algorithms for large $M$.
The parameter $M$ defines the resolution of the initial angular grid which is here given by 
$\Delta \theta^{(0)}=\text{sin}^{-1}(2/M)$. When $M$ is small, the initial grid is coarse and thus the 
algorithms suffer from convergence to local minima resulting in higher channel estimation error. As $M$ increases the grid 
resolution improves which in turn improves the channel estimation performance of the algorithms.
Further, for a fixed number of paths, as $M$ increases the level of sparsity increases and thus the non-zero coefficients are more successfully recovered by the algorithms.

\begin{figure}[tp]
    \begin{minipage}{0.48\textwidth}
        \centering
		
     	\includegraphics[width=0.99\textwidth]{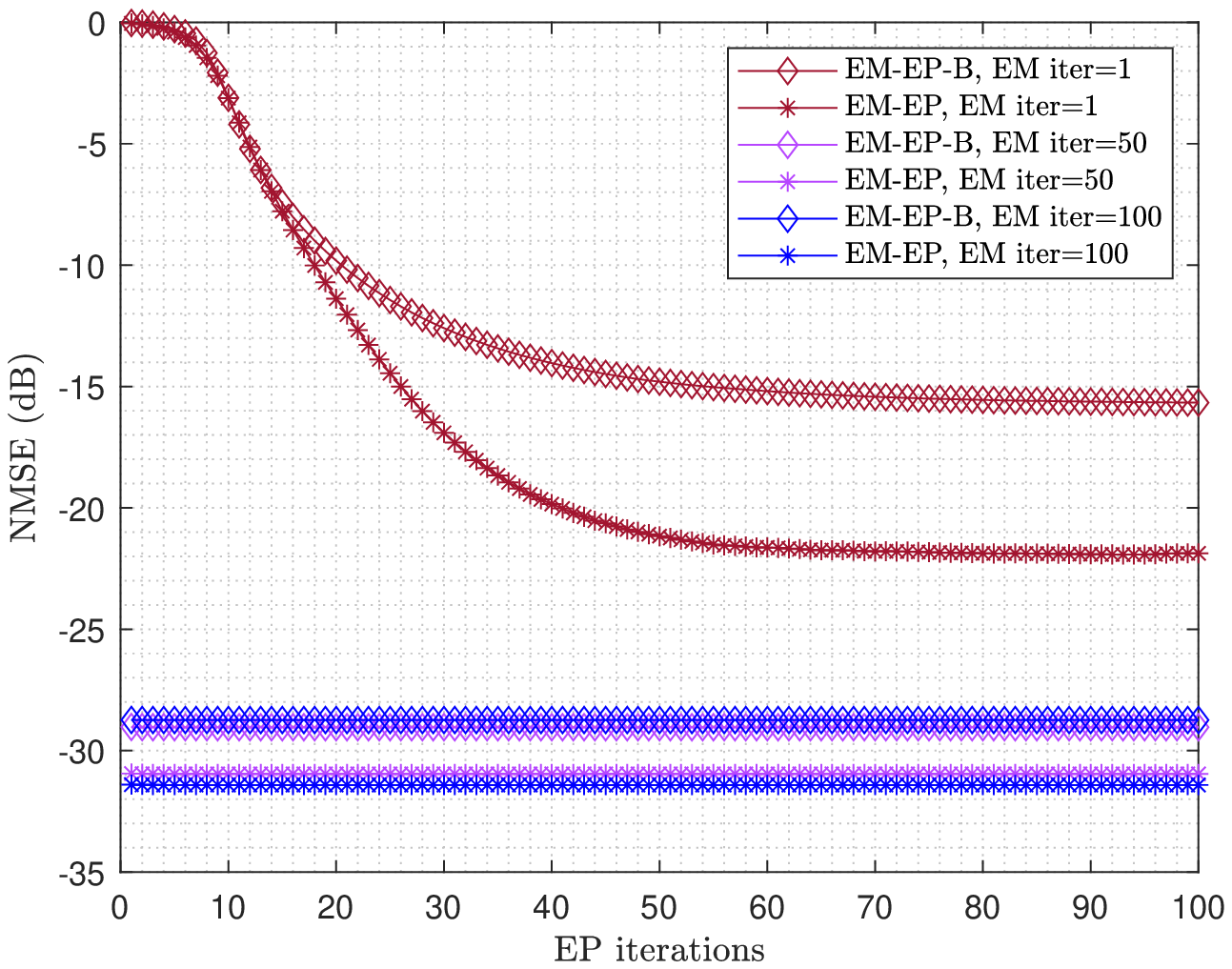}
			\caption{Channel estimation error vs. EP iterations when EM iterations are 
			either 1, 50, or 100, and other parameters are $G=128$, $N=64$, $M=200$, $L_s=3$, $L_p=10$, and $A=10^\text{o}$.}
					\label{fig:NMSE_vs_EP_iter}
    \end{minipage}\hspace{.01\linewidth}
    \begin{minipage}{0.48\textwidth}
        \centering
\includegraphics[width=0.99\textwidth]{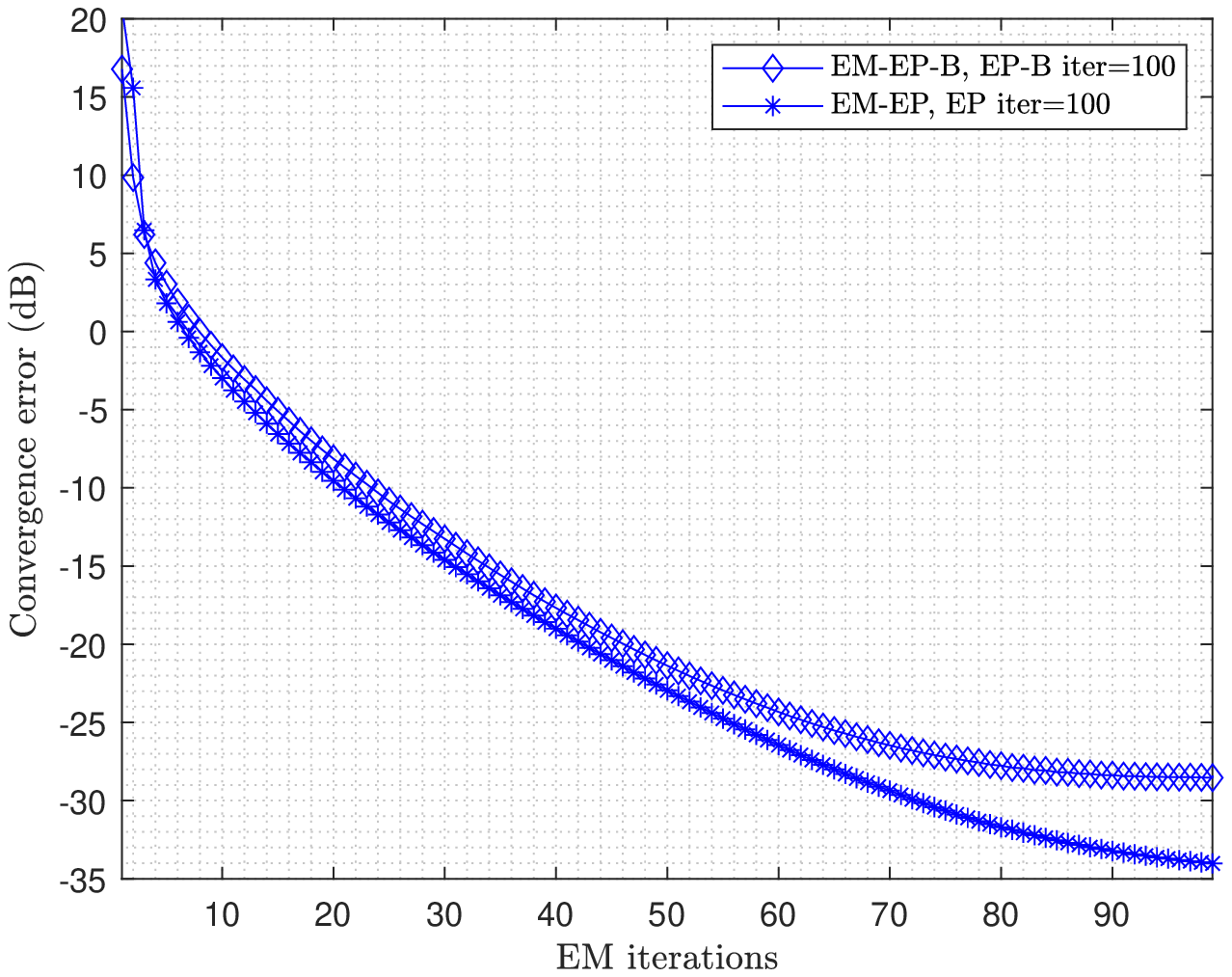}
			\caption{Convergence error of estimating the model parameters 
			$\bdxi$ vs. EM iterations when EP-B and EP iterations are 100, and other parameters are $G=128$, $N=64$, $M=200$, $L_s=3$, $L_p=10$, and $A=10^\text{o}$.}
				\label{fig:Error_vs_EM_iter}
		\end{minipage}
\end{figure}


Fig. \ref{fig:NMSE_vs_EP_iter} shows the average channel estimation performance of the EM-EP-B and EM-EP algorithms 
over the EP iterations when the EM iterations for both algorithms are either fixed to 1, 50, or 100. 
The normalized mean 
squared error plotted in the figure is computed after the selected number of EM iterations are run, and thus the error 
corresponds to the EP run in the last iteration of EM. The estimation error is defined as 
$\text{NMSE (dB)}=10\log_{10} \frac{\Exp \left[||\bdh_{n}-\bdh||^2\right]}{\Exp \left[||\bdh||^2\right]}$ where $\bdh_{n}$ 
is the channel estimate obtained at the $n$-th iteration of the EP-B or EP algorithm. We see that when the EM iteration is fixed to 1, 
our proposed EP algorithm converges faster than the EP-B algorithm with 
a significant improvement in channel estimation. As the EM 
iterations increase both algorithms converge in just 3 iterations, but as expected EM-EP continues to maintain 
a small edge in channel estimation performance. 
This improvement in the convergence performance of the EP-based algorithms is achieved due to the 
following two reasons. 
Each full run of the EP-B and EP algorithms in an EM iteration is initialized using 
the approximate distribution $Q(\bdw,\bdz)$ 
obtained in the previous EM iteration. 
Our experiments show that such initialization of the EP algorithm
results in 
an improved estimation and convergence performances. 
Another reason is that as the EM iterations continue, 
the EM estimates of the model parameters $\bdxi$ converges to 
the local maximum of the likelihood function $p(\bdy|\bdxi)$, 
and the EP-based algorithms closely approximate the true joint 
posterior distribution on the sparse vector and its support.

Finally, in Fig. \ref{fig:Error_vs_EM_iter} we show the average convergence error performance of the EM-EP-B and EM-EP algorithms in estimating the model parameters $\bdxi$ over 
the EM iterations. The iterations of the EP-B and EP algorithms are fixed to 100 and the results are averaged over $1000$ Monte Carlo trials. 
The convergence error plotted in the figure is defined as 
$\text{convergence error (dB)}=10 \log_{10} \frac{||\bdxi^{(l+1)}-\bdxi^l||^2}{||\bdxi^l||^2}$ where index $l$ represents the $l$-th iteration of EM, and $l=1,2,\ldots,n_{EM}$. 
It is observed that in the first four iterations the error of EM-EP is larger than EM-EP-B due to the fact that EM-EP is also estimating the Markov transition probabilities 
$\tau_{01}$ and $\tau_{10}$. 
However, after that EM-EP outperforms EM-EP-B. Moreover, 
after 70-80 iterations EM-EP-B has reached a plateau whereas EM-EP continues to improve.
As observed the convergence error reduces with the EM iterations until the EM estimates converges to a local maximum of the objective 
function. Furthermore, comparing Figs. \ref{fig:NMSE_vs_EP_iter} and \ref{fig:Error_vs_EM_iter}, we observe that after 50 EM iterations, any further EM 
iteration does not improve the channel estimation performance of EM-EP-B algorithm, 
whereas as expected our proposed EM-EP algorithm continues to improve.

\section{Conclusions}\label{conclude_sectn}

We consider the problem of downlink channel 
estimation in the multi-user massive MIMO systems. 
To capture the clustered sparse 
nature of the channel, we assume a conditionally independent 
identically distributed Bernoulli-Gaussian prior 
on the sparse vector representing the channel, and a Markov prior on its support vector. We develop an expectation propagation (EP) based 
algorithm to approximate the intractable joint distribution on the sparse vector and its 
support with a distribution from an exponential family. To find the 
maximum likelihood estimates of the 
hyperparameters and the angular grid points, we integrated the EP algorithm with the expectation 
maximization (EM) algorithm. The resulting EM-EP algorithm directly estimates 
the hyperparameters and the clustered sparse downlink channel. Simulation 
results show that due to the inclusion of the correlations in the sparse vector 
in the approximated posterior and the use of EP method, our EM-EP algorithm can recover the channel 
with non-equal size burst sparsity. Further, 
the proposed EM-EP algorithm outperforms the existing algorithms in the literature
including 
{S-TCS and SuRe-CSBL
algorithms which also use a Markov prior on the support vector.} 

\section*{Appendix A: Proof of Lemma \ref{lemma1}}
Given the hybrid posterior distribution as 
\begin{align}\label{App_R}
R_{2,m}(w_m,z_m)&=\frac{1}{C_m}p(w_m|z_m) \CN(w_m;\mu_{\backslash 2,m},\Sigma_{\backslash 2,m})\times\nonumber \\
&\quad \Bern(z_m;\sigma(p_{\backslash 2,m})),
\end{align}
where the normalization constant $C_m$ is written as 
\begin{align}
C_m&=\sum_{z_m\in\{0,1\}}\int p(w_m|z_m)\CN(w_m;\mu_{\backslash 2,m},\Sigma_{\backslash 2,m})\times\nonumber \\
&\quad \Bern(z_m;\sigma(p_{\backslash 2,m}))dw_m,
\end{align}
First, we compute $\frac{\partial \ln C_m}{\partial \mu^*_{\backslash 2,m}}$ in \eqref{partial_mu1a} which can be written as 
\begin{align}
\frac{\partial \ln C_m}{\partial \mu^*_{\backslash 2,m}}
&=\frac{\Exp_{R_{2,m}}[w_m]}{\Sigma_{\backslash 2,m}}-
\frac{\mu_{\backslash 2,m}}{\Sigma_{\backslash 2,m}},\label{partial_mu1b}
\end{align}
\begin{figure*}[ht]
\normalsize
\setcounter{mytempeqncnt}{\value{equation}}
\setcounter{equation}{\value{mytempeqncnt}}
\begin{align}
&\partial \ln C_m=\frac{1}{C_m}\sum_{z_m\in\{0,1\}}\int p(w_m|z_m)\partial\left[\CN(w_m;\mu_{\backslash 2,m},\Sigma_{\backslash 2,m})\right]
\Bern(z_m;\sigma(p_{\backslash 2,m}))
dw_m,\nonumber\\
&=\frac{1}{C_m}\sum_{z_m\in\{0,1\}}\int p(w_m|z_m)\CN(w_m;\mu_{\backslash 2,m},\Sigma_{\backslash 2,m})
\Bern(z_m;\sigma(p_{\backslash 2,m}))
\left[\frac{w_m-\mu_{\backslash 2,m}}{\Sigma_{\backslash 2,m}}\right]dw_m\partial \mu^*_{\backslash 2,m},\label{partial_mu1a}
\end{align}
\begin{align}
&\partial \ln C_m=
\frac{1}{C_m}\sum_{z_m\in\{0,1\}}\int p(w_m|z_m)\partial \left[\CN(w_m;\mu_{\backslash 2,m},\Sigma_{\backslash 2,m})\right]
\Bern(z_m;\sigma(p_{\backslash 2,m}))dw_m\nonumber\\
&=
\frac{1}{C_m}\sum_{z_m\in\{0,1\}}\int p(w_m|z_m)\CN(w_m;\mu_{\backslash 2,m},\Sigma_{\backslash 2,m})
\Bern(z_m;\sigma(p_{\backslash 2,m})) \nonumber \\
&~~~~~~~~~~~~~~~~~~~ \times \Big[\frac{|w_m-\mu_{\backslash 2,m}|^2}{\left(\Sigma_{\backslash 2,m}\right)^2}
-\frac{1}{\Sigma_{\backslash 2,m}}\Big]dw_m\partial \Sigma_{\backslash 2,m},\label{partial_Sig1a}
\\
&\partial \ln C_m=\frac{1}{C_m}\sum_{z_m\in\{0,1\}}\int p(w_m|z_m)\CN(w_m;\mu_{\backslash 2,m},\Sigma_{\backslash 2,m})
\partial \left[\Bern(z_m;\sigma(p_{\backslash 2,m}))\right]d_{w_m}\nonumber\\
&=\frac{1}{C_m}\sum_{z_m\in\{0,1\}}\int p(w_m|z_m)\CN(w_m;\mu_{\backslash 2,m},\Sigma_{\backslash 2,m})\Bern(z_m;\sigma(p_{\backslash 2,m})) \nonumber \\
&~~~~~~~~~~~~~~~~~~~ \times
\Big[\frac{z_m}{\sigma(p_{\backslash 2,m})}-\frac{(1-z_m)}{(1-\sigma(p_{\backslash 2,m}))}\Big]d{w_m}\partial \sigma(p_{\backslash 2,m}),\label{partial_sig1a}
\end{align}
\begin{align}
&S_{3,m-1,m}(z_{m-1},z_m)\nonumber\\
&\propto \exp\Big\{z_{m-1}\ln\frac{\sigma\left(p^{\backslash R}_{3,m-1}\right)\tau_{10}}{\sigma\left(-p^{\backslash R}_{3,m-1}\right)(1-\tau_{01})}+z_m \ln \frac{\sigma\left(p^{\backslash F}_{3,m}\right)\tau_{01}}{\sigma\left(-p^{\backslash F}_{3,m}\right)(1-\tau_{01})}
\nonumber \\
&~~~~~~~~~~~~~~~~~~~~~~~
+z_{m-1}z_m \ln \frac{(1-\tau_{10})(1-\tau_{01})}{\tau_{10}\tau_{01}}\Big\},\label{App_hybrid_zm_b}
\end{align}
\hrulefill
\end{figure*}
Setting 
$\mu_m=\Exp_{R_{2,m}}[w_m]$ in \eqref{partial_mu1b} and rearranging it we get
\begin{align}\label{App_mu_m}
\mu_m=\mu_{\backslash 2,m}+\Sigma_{\backslash 2,m}\frac{\partial \ln C_m}{\partial \mu^*_{\backslash 2,m}},
\end{align}
Next we compute $\frac{\partial \ln C_m}{\partial \Sigma_{\backslash 2,m}}$ in 
\eqref{partial_Sig1a} and write it as 
\begin{align}
\frac{\partial \ln C_m}{\partial \Sigma_{\backslash 2,m}}&=
\frac{\Exp_{R_{2,m}}\left[\mid w_m-\mu_{\backslash 2,m}\mid^2\right]}{\left(\Sigma_{\backslash 2,m}\right)^2}-\frac{1}{\Sigma_{\backslash 2,m}},\label{partial_Sig1b}
\end{align}
Expanding the $\Exp_{R_{2,m}}[.]$ 
operator in \eqref{partial_Sig1b} and using \eqref{App_mu_m} in it then rearranging gives
\begin{align}
&\Exp_{R_{2,m}}[|w_m|^2]=\Sigma_{\backslash 2,m}+\left(\Sigma_{\backslash 2,m}\right)^2
\frac{\partial \ln C_m}{\partial \Sigma_{\backslash 2,m}}
+|\mu_{\backslash 2,m}|^2+\nonumber\\
&\Sigma_{\backslash 2,m}\mu_{\backslash 2,m}
\frac{\partial \ln C_m}{\partial \mu_{\backslash 2,m}}+\Sigma_{\backslash 2,m}\mu^*_{\backslash 2,m}
\frac{\partial \ln C_m}{\partial \mu^*_{\backslash 2,m}},\label{Sig1a}
\end{align} 
subtracting $|\Exp_{R_{2,m}}[w_m]|^2$ from both sides of \eqref{Sig1a} and using 
\eqref{varRw} and \eqref{App_mu_m} we get
\begin{equation}\label{App_var_m}
\Sigma_{m,m}=\Sigma_{\backslash 2,m}+\left(\Sigma_{\backslash 2,m}\right)^2
\left[\frac{\partial \ln C_m}{\partial \Sigma_{\backslash 2,m}}-
\frac{\partial \ln C_m}{\partial \mu^*_{\backslash 2,m}}\frac{\partial \ln C_m}{\partial \mu_{\backslash 2,m}}\right],
\end{equation}
Finally we compute $\frac{\partial \ln C_m}{\partial \sigma(p_{\backslash 2,m})}$ in 
\eqref{partial_sig1a} which can be written as
\begin{equation}
\frac{\partial \ln C_m}{\partial \sigma(p_{\backslash 2,m})}=
\frac{\Exp_{R_{2,m}}[z_m]}{\sigma(p_{\backslash 2,m})}-\frac{(1-\Exp_{R_{2,m}}[z_m])}{(1-\sigma(p_{\backslash 2,m}))},\label{partial_sig1b}
\end{equation}
rearranging \eqref{partial_sig1b} and using \eqref{meanRz} we get
\begin{equation}
\sigma(p_m)=\sigma(p_{\backslash 2,m})+\sigma(p_{\backslash 2,m})
(1-\sigma(p_{\backslash 2,m}))\frac{\partial \ln C_m}{\partial \sigma(p_{\backslash 2,m})},\label{sig_pm_eq}
\end{equation}
where using \eqref{norm_const}, we compute
\begin{align}
\frac{\partial \ln C_m}{\partial \sigma(p_{\backslash 2,m})}
&=\frac{1}{C_m}\left[\CN(0;\mu_{\backslash 2,m},\Sigma_{\backslash 2,m}+\gamma^{-1}_m)-\right. \nonumber\\
&\quad \left. \CN(0;\mu_{\backslash 2,m},\Sigma_{\backslash 2,m})\right],\label{part_C_m}
\end{align}
inserting \eqref{part_C_m} in \eqref{sig_pm_eq} and again using \eqref{norm_const} gives
\eqref{lem1_sig_Eq}.

\section*{Appendix B: deriving the marginals in \eqref{S3z_a} and \eqref{S3z_b}}
Let the joint probability mass function (pmf) on $z_{m-1}$ and $z_m$ can be 
defined as $p(z_{m-1}=i,z_m=j)=\phi_{ij}$ for $i,j\in\{00,01,10,11\}$. This 
pmf can be written as 
\begin{align}
p(z_{m-1},z_m)
&=\left[\phi_{11}\right]^{z_{m-1}z_m}\left[\phi_{01}\right]^{(1-z_{m-1})z_m}\times\nonumber\\
& \left[\phi_{10}\right]^{z_{m-1}(1-z_m)}\left[\phi_{00}\right]^{(1-z_{m-1})(1-z_m)},\label{pmfz_gen}\\
&\propto \exp\{z_{m-1}\l_1+z_m\l_2+z_{m-1}z_m\l_3\},\label{App_B_joint_pmfz}
\end{align}
where we define
\begin{align}
\l_1=\ln \frac{\phi_{10}}{\phi_{00}}, \quad
\l_2=\ln \frac{\phi_{01}}{\phi_{00}},\quad
\l_3=\ln \frac{\phi_{00}\phi_{11}}{\phi_{01}\phi_{10}},\label{App_l123}
\end{align}
Next we use \eqref{App_l123} and $\sum_{i,j}\phi_{ij}=1$ to get the solution to this system of equations as 
\begin{align}
\phi_{00}=\frac{1}{1+\exp\{\l_1\}+\exp\{\l_2\}+\exp\{\l_1+\l_2+\l_3\}},\label{phi_00}\\
\phi_{01}=\frac{\exp\{\l_2\}}{1+\exp\{\l_1\}+\exp\{\l_2\}+\exp\{\l_1+\l_2+\l_3\}},\label{phi_01}\\
\phi_{10}=\frac{\exp\{\l_1\}}{1+\exp\{\l_1\}+\exp\{\l_2\}+\exp\{\l_1+\l_2+\l_3\}},\label{phi_10}\\
\phi_{11}=\frac{\exp\{\l_1+\l_2+\l_3\}}{1+\exp\{\l_1\}+\exp\{\l_2\}+\exp\{\l_1+\l_2+\l_3\}},\label{phi_11}
\end{align}

Now the joint distribution on $z_{m-1}$ and $z_m$ in our case is given in 
\eqref{hybrid_zm} as
\begin{align}\label{App_hybrid_zm_a}
S_{3,m-1,m}(z_{m-1},z_m)&=q^{\backslash R}_{3,m-1}(z_{m-1})p(z_m|z_{m-1})
q^{\backslash F}_{3,m}(z_{m}),
\end{align}
using \eqref{pr_z}, \eqref{qbR3z}, and \eqref{qbF3z} in \eqref{App_hybrid_zm_a} and simplifying we get \eqref{App_hybrid_zm_b}. Comparing \eqref{App_B_joint_pmfz} and \eqref{App_hybrid_zm_b}, we see that
\begin{align}
\l_1&=\ln\frac{\sigma\left(p^{\backslash R}_{3,m-1}\right)\tau_{10}}{\sigma\left(-p^{\backslash R}_{3,m-1}\right)(1-\tau_{01})},\\
\l_2&=\ln \frac{\sigma\left(p^{\backslash F}_{3,m}\right)\tau_{01}}{\sigma\left(-p^{\backslash F}_{3,m}\right)(1-\tau_{01})},\\
\l_3&=\ln \frac{(1-\tau_{10})(1-\tau_{01})}{\tau_{10}\tau_{01}},
\end{align}
and using the above equations in \eqref{phi_00}-\eqref{phi_11} we get
\begin{align}
\phi_{00}&=\frac{1}{D_m}\sigma(-p^{\backslash R}_{3,m-1})\sigma(-p^{\backslash F}_{3,m})(1-\tau_{01}),\label{EP_phi_00}\\
\phi_{01}&=\frac{1}{D_m}\sigma(-p^{\backslash R}_{3,m-1})\sigma(p^{\backslash F}_{3,m})\tau_{01},\\
\phi_{10}&=\frac{1}{D_m}\sigma(p^{\backslash R}_{3,m-1})\sigma(-p^{\backslash F}_{3,m})\tau_{10},\\
\phi_{11}&=\frac{1}{D_m}\sigma(p^{\backslash R}_{3,m-1})\sigma(p^{\backslash F}_{3,m})(1-\tau_{10})\label{EP_phi_11},
\end{align}
where the normalization constant $D_m$ is given by
\begin{align}\label{App_norm_const_Dm}
&D_m\nonumber\\
&=\sigma(-p^{\backslash R}_{3,m-1})\sigma(-p^{\backslash F}_{3,m})(1-\tau_{01})
+\sigma(p^{\backslash R}_{3,m-1})\sigma(-p^{\backslash F}_{3,m})\tau_{10}
\nonumber\\
& +\sigma(-p^{\backslash R}_{3,m-1})\sigma(p^{\backslash F}_{3,m})\tau_{01}+
\sigma(p^{\backslash R}_{3,m-1})\sigma(p^{\backslash F}_{3,m})(1-\tau_{10}),
\end{align}
Now once $\phi_{ij}$s' are computed in \eqref{EP_phi_00}-\eqref{EP_phi_11}, the 
marginal distributions on $z_{m-1}$ and $z_m$ can be found from
\begin{align}
S_{3,m-1}(z_{m-1})&=\left[\phi_{10}+\phi_{11}\right]^{z_{m-1}}
\left[\phi_{01}+\phi_{00}\right]^{(1-z_{m-1})},\label{marg_S_3z_a}\\
S_{3,m}(z_m)&=\left[\phi_{01}+\phi_{11}\right]^{z_{m}}
\left[\phi_{10}+\phi_{00}\right]^{(1-z_{m})}\label{marg_S_3z_b},
\end{align}
where \eqref{marg_S_3z_a} and \eqref{marg_S_3z_b} is derived from 
\eqref{pmfz_gen} by marginalizing over the other variable. Notice that the 
means of these marginal distributions are given by $\Exp_{S_{3,m-1}}[z_{m-1}]=\phi_{10}+\phi_{11}$ and 
$\Exp_{S_{3,m}}[z_m]=\phi_{01}+\phi_{11}$ which can be easily computed using \eqref{EP_phi_00}-\eqref{EP_phi_11}.

\bibliographystyle{IEEEtran}
\bibliography{References}
\end{document}